\documentclass[12pt,a4paper]{article}

\usepackage{epsfig}
\usepackage{amsmath}
\usepackage{axodraw}
\usepackage{rotating}
\usepackage{multirow}
\usepackage{cite}
\usepackage{mcite}

\usepackage{bold-extra}

\newcommand{\mrm}[1]{\mathrm{#1}}

\newcommand{\tsc}[1]{\textsc{#1}}

\newlength{\tmplen}

\def\lsim{\mathrel{\rlap{\lower4pt\hbox{\hskip1pt$\sim$}}
   \raise1pt\hbox{$<$}}}                
\def\gsim{\mathrel{\rlap{\lower4pt\hbox{\hskip1pt$\sim$}}
   \raise1pt\hbox{$>$}}}                

\newcommand{\alphas}{\alpha_{\mathrm{s}}}

\newcommand{\pT}{\ensuremath{p_{\perp}}}
\newcommand{\pTs}{\ensuremath{p^2_{\perp}}}
\newcommand{\pTsmax}{\ensuremath{p^2_{\perp\mathrm{max}}}}
\newcommand{\kT}{\ensuremath{k_{\perp}}}
\newcommand{\pTmin}{p_{\perp\mathrm{min}}}
\newcommand{\pTo}{p_{\perp 0}}

\newcommand{\GeV}{\ensuremath{\!\ \mathrm{GeV}}}
\newcommand{\TeV}{\ensuremath{\!\ \mathrm{TeV}}}

\renewcommand{\d}{\mathrm{d}}
\newcommand{\e}{\mathrm{e}}

\newcommand{\g}{\mathrm{g}}

\newcommand{\J}{\mathrm{J}}

\newcommand{\p}{\mathrm{p}}
\newcommand{\q}{\mathrm{q}}
\newcommand{\s}{\mathrm{s}}

\renewcommand{\u}{\mathrm{u}}

\newcommand{\Z}{\mathrm{Z}}

\newcommand{\pbar}{\overline{\mathrm{p}}}
\newcommand{\qbar}{\overline{\mathrm{q}}}

\newenvironment{Itemize}{\begin{list}{$\bullet$}%
{\setlength{\topsep}{0.2mm}\setlength{\partopsep}{0.2mm}%
\setlength{\itemsep}{0.2mm}\setlength{\parsep}{0.2mm}}}%
{\end{list}}
\newcounter{enumct}

\newlength{\abstwidth}
\newlength{\captivewidth}
\begin{document}
\sloppy

\pagestyle{empty}

\begin{flushright}
LU TP 09-28\\
MCnet/09/16\\
November 2009
\end{flushright}

\vspace{\fill}

\begin{center}
\renewcommand{\thefootnote}{\fnsymbol{footnote}} 
{\LARGE\bf Multiparton Interactions and Rescattering}%
\footnote{Work supported by the Marie Curie Early Stage Training program
``HEP-EST'' (contract number \makebox{MEST-CT-2005-019626}) and in part by
the Marie Curie RTN ``MCnet'' (contract number
\makebox{MRTN-CT-2006-035606})}\\[10mm]
\renewcommand{\thefootnote}{\arabic{footnote}}
\setcounter{footnote}{0}
{\Large R.~Corke\footnote{richard.corke@thep.lu.se} and %
T.~Sj\"ostrand\footnote{torbjorn@thep.lu.se}} \\[3mm]
{\it Department of Theoretical Physics,}\\[1mm]
{\it Lund University,}\\[1mm]
{\it S\"olvegatan 14A,}\\[1mm]
{\it S-223 62 Lund, Sweden}
\end{center}

\vspace{\fill}

\begin{center}
{\bf Abstract}\\[2ex]
\begin{minipage}{\abstwidth}
The concept of multiple partonic interactions in hadronic events is vital 
for the understanding of both minimum-bias and underlying-event physics. 
The area is rather little studied, however, and current models offer a 
far from complete coverage, even of the effects we know ought to be there.
In this article we address one such topic, namely that of rescattering, 
where an already scattered parton is allowed to take part in another 
subsequent scattering. A framework for rescattering is introduced for the
\textsc{Pythia 8} event generator and fully integrated with normal 
multiparton interactions and initial- and final-state radiation. 
Using this model, the effects on event structure are studied, and 
distributions are shown both for minimum-bias and jet events.
\end{minipage}
\end{center}

\vspace{\fill}

\clearpage
\pagestyle{plain}
\setcounter{page}{1}

\section{Introduction}

The run-up to the start of the Large Hadron Collider (LHC) has led 
to a greatly increased interest in the physics of multiparton 
interactions (MPI) in hadronic collisions (alternatively called
MI for multiple interactions or MPPI for multiple parton--parton
interactions in the literature). The concept of MPI has its roots in ideas
that preceded QCD, and generators based on MPI have been used for decades
to predict LHC physics. It is only in very recent years that the tenets of
MPI have caught on in a broader experimental and theoretical community,
however, as being central aspects for the proper QCD understanding of
collider physics, and not only as handy parameterisations of some unknown
physics. At the same time their role as a potentially limiting factor in
searches and precision measurements has become more apparent.

\begin{figure}
\begin{center}
\begin{picture}(157.5,112.5)(0,0)
\SetScale{0.75}
\SetWidth{4}
\ArrowLine(10,130)(50,130)
\ArrowLine(10,20)(50,20)
\SetWidth{2}
\ArrowLine(50,121)(100,90)
\ArrowLine(50,29)(100,60)
\Gluon(100,60)(100,90){5}{2}
\ArrowLine(100,90)(200,120)
\ArrowLine(100,60)(200,30)
\Gluon(50,127)(150,90){5}{10}
\Gluon(50,23)(150,60){5}{10}
\Gluon(150,60)(150,90){5}{2}
\Gluon(150,90)(200,100){5}{4}
\Gluon(150,60)(200,50){5}{4}
\ArrowLine(50,133)(200,133)
\ArrowLine(50,139)(200,139)
\ArrowLine(50,17)(200,17)
\ArrowLine(50,11)(200,11)
\GOval(50,130)(15,8)(0){0.5}
\GOval(50,20)(15,8)(0){0.5}
\end{picture}\hspace{10mm}
\begin{picture}(157.5,112.5)(0,0)
\SetScale{0.75}
\SetWidth{4}
\ArrowLine(10,130)(50,130)
\ArrowLine(10,20)(50,20)
\SetWidth{2}
\ArrowLine(50,121)(100,90)
\ArrowLine(50,25)(100,60)
\Gluon(100,60)(100,90){5}{2}
\ArrowLine(100,90)(200,120)
\ArrowLine(100,60)(150,60)
\Gluon(50,127)(150,90){5}{10}
\Gluon(150,60)(150,90){5}{2}
\Gluon(150,90)(200,100){5}{4}
\ArrowLine(150,60)(200,40)
\ArrowLine(50,133)(200,133)
\ArrowLine(50,139)(200,139)
\ArrowLine(50,17)(200,17)
\ArrowLine(50,11)(200,11)
\GOval(50,130)(15,8)(0){0.5}
\GOval(50,20)(15,8)(0){0.5}
\end{picture}
\end{center}

\hspace{50mm}(a)
\hspace{57mm}(b)

\caption{(a) Two $2 \rightarrow 2$ scatterings, (b) a $2 \rightarrow 2$
scattering followed by a rescattering
\label{fig:rescattering-schematic}
}
\end{figure}
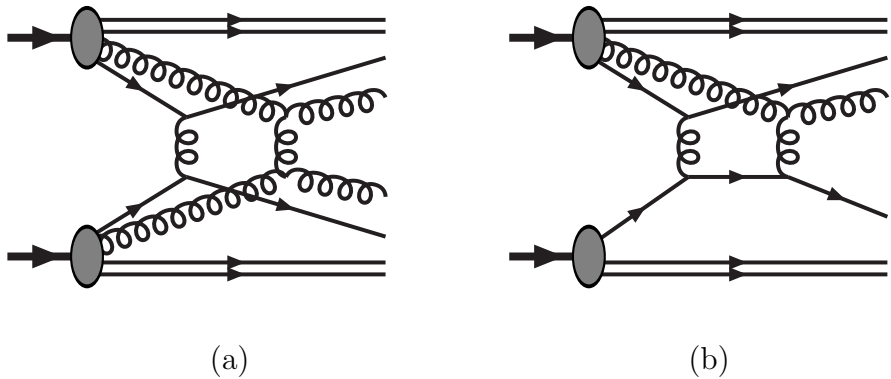

The complexity of MPI physics makes purely analytical studies of
limited validity, and event generators are often the only way to make
contact with data. Existing generators only work within a rather
restricted framework, intended to catch the main aspects, 
but setting aside a number of subleading effects. For instance, 
existing implementations for hadron--hadron collisions are based on 
the assumption of disjoint pairwise scatterings, i.e.\ multiple 
$2 \to 2$ processes, Fig.~\ref{fig:rescattering-schematic}a.
The objective of the current article is to relax this particular
constraint; to introduce a formalism that describes rescattering, 
i.e. where one parton may undergo successive collisions against 
several other partons, in the simplest case leading to a $3 \to 3$ 
process, Fig.~\ref{fig:rescattering-schematic}b. This formalism will 
be integrated into a framework for MPI, initial-state radiation 
(ISR) and final-state radiation (FSR), where everything can be viewed 
as a part of a combined evolution process. 

Before presenting this complex framework, however, it is necessary to 
introduce the various building blocks. In Section~\ref{sec:MPIintro}
the main physics principles of MPI are described, and the current
experimental status summarised. In Section~\ref{sec:MPIinPythia8} we
zoom in on the existing MPI model in \tsc{Pythia 8}, focusing on the 
aspects relevant for the modeling of rescattering.
In Section~\ref{sec:rescattering} rescattering is discussed further, 
and the details of the new implementation in \tsc{Pythia 8} are given. 
Results for a range of different observables are given in 
Section~\ref{sec:results}, before a summary and outlook is presented 
in Section~\ref{sec:conclusions}.

\section{Multiparton Interaction Physics}
\label{sec:MPIintro}

LHC events will typically contain hundreds of outgoing particles.
Some events will contain high-$\pT$ objects --- partons, photons, 
leptons, neutrinos, \ldots --- suggestive of a hard-scattering process 
at large virtuality, while others will not, i.e.\ all particle production
occurs at small $\pT$ scales, say around or below 1 GeV. The dividing 
line is not clearcut, however. As the lower cutoff $\pTmin$
is reduced, the jet rate continuously increases, until eventually
all particles in all events belong to a jet. Well before this stage, 
the direct relationship of a jet to a high-$\pT$ object is gone. 

Nevertheless, there is a problem; perturbative QCD not only predicts
a large number of jets per event, but an infinite number. The diverging 
jet cross section is readily visible in the cross section for $2 \to 2$ 
QCD processes
\begin{equation}
\frac{\d\sigma}{\d\pTs} = \sum_{i,j} \int \d x_1 \int \d x_2 \,
f_i(x_1, Q^2) \, f_j(x_2, Q^2) \, \frac{\d \hat{\sigma}}{\d\pTs} ~,
\label{eq:dsigmadpt}
\end{equation} 
where $\d \hat{\sigma} / \d\pTs$ contains a $1/\pT^4$ divergence from 
$t$-channel gluon exchange (Rutherford scattering), further steepened 
by the rise of the parton densities $f_i, f_j$ at small $x$. Part of
the solution should be that perturbation theory must break down when
$\pT$ becomes of order $\Lambda_{\mrm{QCD}}$, or equivalently of inverse 
hadronic size. Another part is to note that eq.~(\ref{eq:dsigmadpt})
is an inclusive cross section which does not tell how scatterings are 
distributed inside physical events. The task of an MPI model is 
therefore to regularise the divergence at $\pT \to 0$ by nonperturbative 
effects and recast the remaining jet cross section into a probability for 
an event to contain one or several $2 \to 2$ subprocesses in various 
configurations.   

Historically, MPI was originally studied in two different contexts
(see \cite{Sjostrand:2004pf} for a longer survey with references): 
\begin{enumerate}
\item By the exchange of multiple Pomerons. Translated into 
QCD language, a cut Pomeron can be viewed as the transfer of a 
zero-momentum gluon between the two colliding hadrons, transforming 
them into coloured outgoing beam remnants. Colour fields are stretched
longitudinally in between, and eventually fragment to give 
low-$\pT$ particles. Thus, one obtains a model e.g. for multiplicity 
and rapidity distributions, but without any jets.
\item By the perturbative calculation of two or more (semi-)hard 
collisions, thus contributing to the rate of multijets with 
specific jet correlation signatures, but without any underlying events.
\end{enumerate}
These two approaches first came together in a model \cite{Sjostrand:1987su}
that extended the perturbative picture to such low $\pT$ scales that
one could view all events as containing one or more interactions.
These interactions would fill the function of cut Pomerons, giving rise
to colour fields that stretch across the event and hadronise to give
the observable final state. The twist is that now these fields could 
also stretch out to partons of varying $\pT$, thereby giving a smooth
transition to (mini)jets, and a unified picture of minimum-bias (MB) 
and underlying-event (UE) physics.
The main parameter of the model is a ``colour-screening'' $\pTo$ scale 
that is used to dampen the QCD $1/\pT^4$ divergence to 
$1/(\pTo^2 + \pTs)^2$. The modeling also involves varying impact 
parameters (with central collisions likely to involve more activity 
than peripheral ones), corrected parton densities (to ensure momentum 
conservation), a possibility of colour reconnection (to ensure a rising 
$\langle \pT \rangle (n_{\mrm{charged}})$), and a few other aspects.
It was implemented in the \textsc{Pythia} event generator 
\cite{Bengtsson:1987kr,Sjostrand:2006za}. 

For a number of years the model remained unchanged, but in recent years
it has been improved. For instance, the original model only treated the
highest $\pT$ interaction in full, limiting subsequent 
ones to give $\q\qbar$ or $\g\g$ final states without any showers. With 
the introduction of junction fragmentation it became possible to handle 
more complicated beam remnants and thus to consider arbitrary colliding 
flavour combinations \cite{Sjostrand:2004pf}, including showers for all 
interactions. The subsequent introduction of $\pT$-ordered showers
\cite{Sjostrand:2004ef}, instead of the older virtuality-ordered ones, 
allowed a common interleaved evolution of MPI and ISR downwards in $\pT$, 
one of the points being to take into account the competition for 
beam momentum between these two components. An annealing algorithm for
colour reconnection among final-state partons 
\cite{Sandhoff:2005jh,*Skands:2007zg} complements the picture of the most
sophisticated versions available in \textsc{Pythia 6.4}, but also the 
older options are available for backwards compatibility. 

The C++-based \textsc{Pythia 8.1} \cite{Sjostrand:2007gs} does away 
with many older options. It retains the newer beam-remnant handling and
the $\pT$-ordered showers, now also with FSR interleaved in 
the common $\pT$-ordered evolution sequence by default. It allows for a broader 
range of $2 \to 2$ processes to be included in the MPI framework, with 
outgoing photons, leptons, $\J/\psi$ and $\Upsilon$ possible, to allow
a more realistic mix of possible background processes. Another new feature
is the ability to select the nature not only of the hardest subcollision in
an event but also of the second hardest. The current colour reconnection
algorithm is less flexible than the annealing one in \textsc{Pythia 6.4},
however. More details on the \textsc{Pythia 8.1} framework are found
in the subsequent sections.

The \textsc{Phojet} generator \cite{Engel:1994vs,*Engel:1995yda} has its
historical roots in the cut Pomeron approach, with the original soft 
interactions complemented with a hard $2 \to 2$ component.
Thus, an unmodified perturbative component is applied down to 
a cutoff scale, $\pTmin$, while below this, the soft nonperturbative 
component takes over. In practice the difference to \textsc{Pythia}'s 
single dampened perturbative component is minor, since the 
region around or below $\pTmin/\pTo$ does not give rise to visible jets.
\textsc{Phojet} has an eikonal description of the impact-parameter picture
and a component that allows diffraction to be handled as an integrated part
of the MPI framework.

One of the simplest ways of simulating the MB and UE activity is through 
a parameterisation of the main aspects of the data, such as that of 
UA5 \cite{Alner:1986is}. The emphasis here is on the charged multiplicity
distribution and its pseudorapidity distribution. This model was later
implemented in the \textsc{Herwig} event generator 
\cite{Marchesini:1988hj,*Marchesini:1991ch}. 
Although such parameterisations can be tuned to describe some 
distributions well, they are not reliable when extrapolated beyond 
the fit region, and fail to describe correlations and fluctuations. 
The \textsc{Jimmy} add-on \cite{Butterworth:1996zw} for \textsc{Herwig} 
uses an eikonal model and shares many features both with \textsc{Phojet} 
and with the original \textsc{Pythia} model. It is only a model for
underlying events, however. The model was further extended to include 
soft interactions in \cite{Borozan:2002fk}, and it is this latter model
that forms the basis of the MPI model in \textsc{Herwig++} 
\cite{Bahr:2008dy,*Bahr:2008pv}. 

The most obvious prediction of the MPI approach is that there 
should be a subsample of events that contains two separate $2 \to 2$ 
subcollisions, known as double parton scattering (DPS). Experimentally, the 
problem is that a four-jet sample can contain a significant fraction 
of $2 \to 4$ events, so-called double bremsstrahlung (DBS). The 
fall-off of the MPI cross section is steeper than that of 
bremsstrahlung, so that DPS is dominant only at rather small 
$\pT$ scales. The trick is therefore to do the analysis with as low 
$\pTmin$ thresholds as possible, while still keeping the jets 
sufficiently hard that the jet momentum can reasonably well be 
associated with that of an originating parton. In this way one can 
study kinematical distributions that distinguish DPS and DBS events,
such as that the former ought to consist of two parton pairs lined up
back-to-back in azimuth. The first measurement of this kind was performed
by the AFS collaboration at ISR \cite{Akesson:1986iv}. So far, the clearest
observation has been provided by the CDF collaboration in a study of
$\gamma+3$ jet events \cite{Abe:1997xk}. Recently a preliminary D0 study
\cite{D0:2009zz} of the same channel confirms the CDF results with even
more statistics.

More indirect evidence comes from a number of other observables, 
starting from the fact that observed multiplicity distributions 
are much broader than can easily be understood in terms of only
one interaction per event, but are well explained by a varying number
of MPI. This variation can also explain the observed strong 
forward--backward correlations in multiplicity. 

The CDF studies on MB and UE activity
\cite{Field:2002vt,Field:2005sa,Field:2006iy,Kar:2009kc,Field:2009zz} are
also well known. These study, in detail, the old UA1 ``pedestal effect''
observation \cite{Albajar:1988tt} that the activity in events triggered by
the presence of a high-$\pT$ jet is much higher, also well away from the
jets, than in the bulk of normal events. The rise of this activity is very
steep up to a jet $\pT$ of roughly 10 GeV, and thereafter rises only very
slowly. The latter rise can be attributed to contributions from ISR and
FSR, while the sudden change of behaviour at around 10 GeV is a feature of
MPI models with an impact-parameter dependence.

Since not all aspects of the models can be derived from first principles
there are a number of parameters that have to be tuned to data, such as
those in the MPI, ISR, FSR, beam-remnant and hadronisation frameworks. 
Therefore the tuning and testing of these models go hand in hand.
A well-known example is the tunes of R.D. Field 
\cite{Field:2005sa,Field:2006iy,Field:2009zz}, such as Tune A and
Tune DW, originally intended for the CDF collaboration but with 
widespread usage. These are based on the older \textsc{Pythia}
models, while the Perugia \cite{Skands:2009zm} and Professor 
\cite{Buckley:2009bj} tunes are for the most recent \textsc{Pythia 6.4} 
models. Other tunes have been developed inside experimental collaborations
\cite{Moraes:2007rq}.

These tunes are used to try to get an insight into
what can be expected at new experiments, specifically at the LHC. Such
extrapolations, however, come with a high level of uncertainty; within many
of the models in a generator there are parameters with an unknown energy
scaling. There is, therefore, also the exciting prospect that new data will
further constrain and improve models, as well as revealing new insights.
MPI is one of the least well understood areas. While current models, after
tuning, are able to describe many distributions very well, there are still
many others which are not fully described. This is a clear sign that new
physics aspects need to be introduced, and it is therefore not enough to
``sit still'' while waiting for new data. It is with this in mind that we
look at rescattering, where an already scattered parton is able to undergo
another subsequent scattering.  Such processes have been studied
analytically already many years ago, notably by Paver and Treleani
\cite{Paver:1983hi,*Paver:1984ux}. There, estimates were given for the size of
rescattering effects based on a set of ans\"atze for the calculation of
parton distribution functions (PDFs). As with MPI in general, although
such rescatterings may be relatively soft, they can lead to non-trivial
colour flows which can change the structure of events.

The theoretical study of MPI and rescattering, in various approaches, 
has received renewed attention in recent years \cite{Cattaruzza:2004qr,%
Bartels:2005wa,Khoze:2006uj,Avsar:2006jy,Flensburg:2008ag,Calucci:2009sv}.

The experimental study can be traced at least back to the so-called 
Cronin effect \cite{Cronin:1974zm}. The observation is that in 
proton-nucleus collisions the ``high-$\pT$'' particle production is 
enhanced and the low-$\pT$ one reduced, relative to a simple (geometrical) 
rescaling of the proton-nucleon production rates. The interpretation is 
not unambiguous, but some form of rescattering is likely to be involved,
whether at the parton or at the hadron level. Rescattering is nowadays
taken for granted in the study of heavy-ion collisions at RHIC, say, for
instance as an important aspect of the initial formation of a quark--gluon
plasma. In this busy environment it is not possible to study the individual
rescattering processes, however, but only to deduce their collective
importance.

To the best of our knowledge, a model for the detailed effects of 
rescattering in hadron--hadron collisions has never before been
published. It is precisely the task of this article to present a 
first such complete model, formulated in the context of the 
\textsc{Pythia 8} event generator, fully taking into account 
all other aspects of complete events, including hard processes,
initial- and final-state radiation, colour flow, and hadronisation.
This program is publicly available. It thus invites experimental tests, 
not restricted by the imagination of the current authors, but by 
that of the experimental community.

\section{Multiparton Interactions in \tsc{Pythia 8}}
\label{sec:MPIinPythia8}

We begin this section by reviewing the existing MPI framework present 
in \tsc{Pythia 8}, with special attention paid to the areas which will 
later be affected by the introduction of rescattering. For a more 
comprehensive discussion, readers are directed 
to \cite{Sjostrand:2006za} and the references therein. There are some
aspects, like the complete interleaving of ISR, FSR and MPI and the 
related dipole structure of radiation, that are new to \tsc{Pythia 8} 
and are presented here for the first time.

The master equation for the new interleaved evolution is 
\begin{eqnarray}
\frac{\d \mathcal{P}}{\d \pT} & = &
\left( \frac{\d \mathcal{P}_{\mrm{MPI}}}{\d \pT} +
\sum \frac{\d \mathcal{P}_{\mrm{ISR}}}{\d \pT} +
\sum \frac{\d \mathcal{P}_{\mrm{FSR}}}{\d \pT}
\right) \nonumber \\
& \times & \exp \left( - \int_{\pT}^{p_{\perp i-1}} \left(
\frac{\d \mathcal{P}_{\mrm{MPI}}}{\d \pT'} +
\sum \frac{\d \mathcal{P}_{\mrm{ISR}}}{\d \pT'} +
\sum \frac{\d \mathcal{P}_{\mrm{FSR}}}{\d \pT'}
\right) \d \pT' \right)
~.
\label{eqn:pTevol}
\end{eqnarray}
Here $\pT = {\pT}_i$ is the transverse momentum of the  $i^{th}$ 
interaction or shower branching to take place, with a combined 
Sudakov-like form factor expressing that nothing happens between
the previous $(i - 1)^{th}$ interaction/branching and the current
one.

The physics content of this equation is as follows. A hard process 
occurs at a scale ${\pT}_{\mrm{hard}} = {\pT}_1$. At a scale 
${\pT}_2 < {\pT}_1$ either there is a second hard interaction, 
one of the incoming partons radiates, or one of the outgoing partons 
radiates. The respective probability for each of these to occur 
at a generic scale $\pT$ is schematically designated  
$\d \mathcal{P}_{\mrm{MPI}} / \d \pT$,
$\d \mathcal{P}_{\mrm{ISR}} / \d \pT$, and
$\d \mathcal{P}_{\mrm{FSR}} / \d \pT$, respectively.
Their sum gives the total probability for something to happen at $\pT$.
The requirement that ${\pT}_2$ is the second hardest scale implies
that nothing happened in the range ${\pT}_1 > \pT > {\pT}_2$.
Assuming that the no-activity probability factorises, 
$\mathcal{P}_{\mrm{no}}({\pT}_1  > \pT > {\pT}_2) =
\mathcal{P}_{\mrm{no}}({\pT}_1  > \pT > {\pT}') \times
\mathcal{P}_{\mrm{no}}({\pT}'  > \pT > {\pT}_2)$,
and conservation of probability then gives the exponential in the above
expression. 

The story now repeats for the next scale, ${\pT}_3 < {\pT}_2$.
If the previous step gave an MPI then there are two systems,
either of which can undergo ISR or FSR. If instead it gave 
an ISR or FSR then the number of final-state partons of that system 
increased by one. Therefore, in either case, the sums in
eq.~(\ref{eqn:pTevol}) run over an increased number of partons.
Furthermore, several partons may have obtained changed momenta, e.g. by
recoil effects from an ISR emission, and the momentum in the beam remnants
may have been reduced and their flavour composition changed. There are
thus several reasons why the probability expressions in
eq.~(\ref{eqn:pTevol}) change at the ${\pT}_2$ scale; the simple notation
of eq.~(\ref{eqn:pTevol}) hides a lot of complexity.

In summary, for each new lower $\pT$ scale there is one more scattering 
subsystem or one more outgoing parton in an existing subsystem. 
The sequence extends until some lower cutoff is reached, which may be 
different for the three components. Below this scale the probability is 
set to zero, and the $\pT$ sequence terminates. A smooth turnoff would
be more realistic than a sharp cutoff, but the principle is the same.

One important aspect of the interleaved evolution is that it introduces 
an element of competition, specifically that the remaining pool of 
beam-remnant momentum is gradually reduced, and therefore the 
phase space for new interactions or emissions shrinks correspondingly.  
One could contrast our approach with two alternatives. In one extreme each 
interaction is allowed to shower in full before the next interaction
is considered. In that case one tends to get fewer interactions, with
each interaction having more activity, especially the first one.  
In the other extreme all interactions are found before showers are added,
which then gives more interactions per event with less radiation for each.
Differences should not be exaggerated, however, since most of the momentum
remains in the remnants for the majority of events.

It would be tempting to view the ordering in terms of decreasing $\pT$
as one of increasing time, in the spirit of the Heisenberg uncertainty 
principle. This would be too simpleminded, however. MPI are supposed
to occur when the two (more or less) Lorentz contracted hadron ``pancakes''
pass through each other, and if so, the different MPI's occur simultaneously
rather than in any causal order. At that time the ISR should already 
``be there'', in terms of virtual fluctuations that are put on the mass
shell by the collisions. Here, time stretches backwards, with low-$\pT$
fluctuations able to live longer and thus be produced earlier than 
high-$\pT$ ones. So it is only for FSR that formation time arguments 
gives the desired answer. Instead we would like to view
eq.~(\ref{eqn:pTevol}) primarily as an expression of conditional
probabilities; \textit{given} that we already know the event structure if
viewed at some resolution scale $\pT$, what further activity is allowed and
with what rate at a lower $\pT$ resolution? In this sense we are
zooming in on an event that is ``already'' there, only resolving finer and
finer details. At larger scales the system is still simple, and so
perturbation theory should provide a good answer.  At lower scales the
event complexity mounts and so further modeling becomes increasingly
difficult, but then again these details will play a lesser role. Of course,
implicit in this is that we already know that $\pT$ ordering is a sensible
choice for ISR, FSR and MPI individually, and thus presumably for the
combination of the three.

A final, technical note on eq.~(\ref{eqn:pTevol}). It would seem to be a 
major task to generate events according to this non-trivial expression
that can sum over many terms, each with a lot of complexity. However, 
there is a simple solution, ``the winner takes it all''. The idea is that 
each term individually is used to find a $\pT$ scale, with Monte Carlo 
methods, and then the one with largest $\pT$ is selected to occur. We 
illustrate this with simplified notation and only two probabilities, 
$\mathcal{P}_a(t)$ and $\mathcal{P}_b(t)$, with $t$ increasing:
\begin{eqnarray}
\mathcal{P}(t) & = & \left( \mathcal{P}_a(t) + \mathcal{P}_b(t) \right)  
\times \exp \left( - \int_{t_{i-1}}^{t} 
\left( \mathcal{P}_a(t') + \mathcal{P}_b(t') \right) \d t' \right)
\nonumber \\ & = &
\left[ \mathcal{P}_a(t) \, \exp \left( - \int_{t_{i-1}}^{t}
\mathcal{P}_a(t')  \d t' \right) \right] \times 
\exp \left( - \int_{t_{i-1}}^{t} \mathcal{P}_b(t')  \d t' \right)
\nonumber \\ & + &
\left[ \mathcal{P}_b(t) \, \exp \left( - \int_{t_{i-1}}^{t}
\mathcal{P}_b(t')  \d t' \right) \right] \times 
\exp \left( - \int_{t_{i-1}}^{t} \mathcal{P}_a(t')  \d t' \right) ~.
\label{eqn:pTwinner}
\end{eqnarray}
Here the expression inside the first square brackets is the probability
to select a $t_a = t_i$ according to $\mathcal{P}_a$ alone, as can 
be done e.g. with the veto algorithm \cite{Sjostrand:2006za}, while the
subsequent factor is the probability that $\mathcal{P}_b$ did \textit{not} 
pick a $t_b$ in the range $t_{i - 1} < t_b < t_a$. That is, that $a$ is
picked and $t_a < t_b$ , while correspondingly the last term is the 
probability that $b$ is picked and $t_b < t_a$. From here it is easy
to generalise to the original task. 

\subsection{Multiparton Interactions}
\label{sec:MPIinPythia8.MPI}

If we now focus on just the contribution from MPI, the probability for an
interaction is given by
\begin{equation}
\frac{\d \mathcal{P}_{\mrm{MPI}}}{\d {\pT}} =
\frac{1}{\sigma_{nd}} \frac{\d \sigma}{\d \pT} \;
\exp \left( - \int_{\pT}^{p_{\perp i-1}}
 \frac{1}{\sigma_{nd}} \frac{\d \sigma}{\d \pT'} \d \pT' \right)
,
\label{eqn:MPIevol}
\end{equation}
where $\sigma_{nd}$ is the non-diffractive inelastic cross section and 
$\d \sigma / \d \pT$ is given by the perturbative QCD $2\rightarrow2$
cross section in non-diffractive events (we leave the issue of 
diffractive events aside for this article).
This cross section is dominated by $t$-channel gluon
exchange, and diverges roughly as $\d \pT^2 / \pT^4$. To avoid this
divergence, the idea of colour screening is introduced. The concept of a
perturbative cross section is based on the assumption of free incoming
states, which is not the case when partons are confined in colour-singlet
hadrons. One therefore expects a colour charge to be screened by the
presence of nearby anti-charges; that is, if the typical charge separation
is $d$, gluons with a transverse wavelength $\sim 1 / \pT > d$ are no
longer able to resolve charges individually, leading to a reduced effective
coupling. This is introduced by reweighting the interaction cross section
such that it is regularised according to
\begin{equation}
\frac{\d \hat{\sigma}}{\d \pT^2} \propto
\frac{\alphas^2(\pT^2)}{\pT^4} \rightarrow
\frac{\alphas^2({\pT^2}_0 + \pT^2)}{({\pT^2}_0 + \pT^2)^2}
,
\label{eqn:pt0}
\end{equation}
where $\pTo$ (related to $1 / d$ above) is now a free parameter in the
model. To be more precise, it is the physical cross section $\d \sigma / \d
\pT^2$ that needs to be regularised, i.e.  the convolution of
$\d \hat{\sigma} / \d \pT^2$ with the two parton densities,
eq.~(\ref{eq:dsigmadpt}). One is thus at liberty to associate the screening
factor with the incoming hadrons, half for each of them, instead of with
the interaction. Such an association also gives a recipe to regularise the
ISR divergence, which goes like $\alphas(\pT^2) / \pT^2$.

Not only $\pTo$ itself, as determined e.g.\ from Tevatron data, 
comes with a large uncertainty, but so does the energy scaling of this 
parameter. The ansatz for the energy dependence of $\pTo$ is that
it scales in a similar manner to the total cross section, 
i.e.\ driven by an effective power related to the Pomeron intercept
\cite{Donnachie:1992ny}, which in turn could be related to the 
small-$x$ behaviour of parton densities. This leads to a scaling
\begin{equation}
\pTo(E_{\mathrm{CM}}) = p_{\perp0}^{\mathrm{ref}} \times \left(
\frac{E_{\mathrm{CM}}}{E_{\mathrm{CM}}^{\mathrm{ref}}}\right)%
^{E_{\mathrm{CM}}^{\mathrm{pow}}} ~,
\label{eqn:pT0scaling}
\end{equation}
where $E_{\mathrm{CM}}^{\mathrm{ref}}$ is some convenient reference energy 
and $p_{\perp0}^{\mathrm{ref}}$ and $E_{\mathrm{CM}}^{\mathrm{pow}}$ are
parameters to be tuned to data. There is no guarantee that this is the
correct shape, however; some studies have suggested that the rise of 
$\pTo$ may flatten out at higher energies 
\cite{Dischler:2000pk,Gustafson:2002kz}. 

Up to this point, all parton-parton interactions have been assumed to be
independent, such that the probability to have $n$ interactions in an event,
$\mathcal{P}_n$, is given by Poissonian statistics. This picture is now
changed, first by requiring that there is at least one interaction, such
that we have a physical event, and second by including an impact parameter,
$b$. For a given matter distribution, $\rho(r)$, the time-integrated
overlap of the incoming hadrons during collision is given by
\begin{equation}
\mathcal{O}(b) =
\int \d t \int \d^3 x \;
\rho(x, y, z) \;
\rho(x + b, y, z + t)
,
\end{equation}
after a suitable scale transformation to compensate for the boosted nature
of the incoming hadrons. 

Such an impact parameter picture has central collisions being generally
more active, with an average activity at a given impact parameter being
proportional to the overlap, $\mathcal{O}(b)$. While requiring at least one
interaction results in $\mathcal{P}_n$ being narrower than Poissonian, when
the impact parameter dependence is added, the overall effect is that
$\mathcal{P}_n$ is broader than Poissonian. The addition of an impact
parameter also leads to a good description of the ``Pedestal Effect'',
where events with a hard scale have a tendency to have more underlying
activity; this is as central collisions have a higher chance both of
a hard interaction and of more underlying activity. This centrality effect
naturally saturates at ${\pT}_{hard} \sim 10 \GeV$ or, more specifically,
when $\sigma_{\mrm{jet}}(\pT > {\pT}_{hard}) \ll \sigma_{\mrm{nd}}$ 
\cite{Sjostrand:1987su}.

ISR, FSR and MPI can all lead to changes in the incoming PDFs. With FSR, a
colour dipole can stretch from a radiating parton to a beam remnant,
leading to (a modest amount of) momentum shuffling between the beam and the
parton. Both ISR and MPI can result in large $x$ values being taken from
the beams, as well as leading to flavour changes in the PDFs.

In the original model, PDFs were rescaled only such that overall momentum
was conserved. This was done by evaluating PDFs at a modified $x$ value
\begin{equation}
x_i' = \frac{x_i}{1 - \sum^{i-1}_{j=1} x_j}, 
\end{equation}
where the subscript $i$ refers to the current interaction and the sum runs
over all previous interactions. The current model still conserves momentum,
but does it in a more flavour-specific manner, as follows. If a valence
quark is taken from one of the incoming hadrons, the valence PDF is
rescaled to the remaining number. If, instead, a sea quark ($\mrm{q_s}$) is
taken from a hadron, an anti-sea companion quark ($\mrm{q_c}$) is left
behind. The $x$ distribution for this companion quark is generated from a
perturbative ansatz, where the sea/anti-sea quarks are assumed to have come
from a gluon splitting, $\mrm{g \rightarrow q_s q_c}$. Subsequent
perturbative evolution of the $\mrm{q_c}$ distribution is neglected. Note
that if a valence quark is removed from a PDF, momentum must be put back
in, while if a companion quark is added, momentum must be taken from the
PDF, in order to conserve momentum overall. This is done by allowing the
normalisation of the sea and gluon PDFs to fluctuate.

\subsection{Initial- and Final-State Radiation}
\label{sec:MPIinPythia8.ISR.FSR}

As this framework now includes radiation from all MPI scattering systems, we
will later need to consider radiation from systems that have an incoming
rescattering parton. For this reason, we take some time to briefly review
the parton shower algorithm, including some details of recoil kinematics.

Both initial and final-state radiation are governed by the DGLAP equations
\cite{Gribov:1972ri,*Altarelli:1977zs,*Dokshitzer:1977sg}
\begin{equation}
\d {\cal P}_a(z, t) =  \frac{\d t}{t} \, \frac{\alphas}{2 \pi} \,
P_{a \to bc}(z) \, \d z ~,
\label{eq:DGLAP}
\end{equation}
giving the probability that a parton $a$ will branch into partons $b$ and
$c$ at some scale $t$, with parton $b$ taking fraction $z$ of the energy
of $a$ and parton $c$ a fraction $(1 - z)$. The splitting kernels $P_{a \to
bc}(z)$ give the universal collinear limit of the matrix-element
expressions for $\q \to \q\g$, $\g \to \g\g$ and $\g \to \q\qbar$
branchings. Within the parton shower, an ordering is introduced through a
Sudakov form factor
\cite{Sudakov:1954sw}
\begin{equation}
{\cal P}^{\mrm{no}}_a(t_{\mrm{max}},t) = \exp \left( -
\int_{t}^{t_{\mrm{max}}} \int_{z_{\mrm{min}}}^{z_{\mrm{max}}}
\d {\cal P}_a(z', {t'}^2)    \right) ~,
\label{eq:Sudakov}
\end{equation}
which unitarises the na\"ive splitting probability given by the DGLAP
equations. Picking the evolution variable $t$ to be the squared transverse
momentum (described below), the overall emission probability is given by
\begin{equation}
\d\mathcal{P}_a = \frac{\d \pTs}{\pTs} \, \frac{\alphas(\pTs)}{2\pi} \,
P_{a \to bc}(z) \, \d z \; \mathcal{P}^{\mrm{no}}_a(\pTsmax,\pTs) ~,
\label{eq:evolution-FSR}
\end{equation}
where we note that the scale choice for $\alphas$ is selected to be $\pTs$
independently of the choice of the evolution variable. This formulation is
already suitable for final-state showers, where outgoing partons may have
timelike virtualities that evolve downwards towards on-shell partons.
For ISR, however, the partons build up increasingly spacelike virtualities
as they approach the interaction point. In this scenario, it is customary 
to use backwards evolution \cite{Sjostrand:1985xi,Bengtsson:1986gz}, where 
an already selected parton $b$ may become unresolved into a parton $a$, 
given by an overall probability
\begin{equation}
\d\mathcal{P}_b = \frac{\d\pTs}{\pTs} \, \frac{\alphas(\pTs)}{2\pi} \,
\frac{x' f_a(x',\pTs)}{x f_b(x,\pTs)} \, P_{a \to bc}(z) \, \d z \;
\mathcal{P}^{\mrm{no}}_b(x,\pTsmax,\pTs) ~,
\label{eq:evolution-ISR}
\end{equation}
where now also the Sudakov form factor involves the ratio of parton
densities.

The $\pTs$ evolution variable can be related to the virtuality $Q^2$
by considering the lightcone kinematics ($p^{\pm} = E \pm p_z$) of a 
parton $a$ moving along the $+z$ axis, splitting into parton $b$ with 
$p_b^+ = z p_a^+$ and parton $c$ with $p_c^+ = (1-z) p_a^+$. 
Conservation of $p_-$ then gives
\begin{equation}
\pTs = z (1-z) m_a^2 - (1-z) m_b^2 - z m_c^2~.
\label{eq:pTlightcone}
\end{equation}
For final state radiation, $Q^2 = m_a^2$ with $m_b = m_c = 0$, while for
initial state radiation, $Q^2 = -m_b^2$ with $m_a = m_c = 0$, giving
\arraycolsep 0.7mm
\begin{eqnarray}
\mrm{FSR:} ~~ \pTs &=& z(1-z)Q^2~, \nonumber \\ 
\mrm{ISR:} ~~ \pTs &=& (1-z)Q^2~.
\label{eq:pTlightconeFSR}
\end{eqnarray}
The detailed kinematics, to be described below, does not use a lightcone
$z$ definition, but rather an energy fraction in a specified reference
frame. The $Q^2$ extracted from eq.~(\ref{eq:pTlightconeFSR}), together
with the new $z$ interpretation, therefore gives a true $\pTs$ slightly
smaller than the one above. 

An important new feature of the \pTs-ordered showers is the approach taken
to recoil. This approach is inspired by a cascade formulated in terms of
radiation from dipoles, which has been implemented in the \tsc{Ariadne}
program \cite{Gustafson:1986db,*Gustafson:1987rq,*Lonnblad:1992tz}. In this
hybrid-dipole approach, each radiating parton has associated with it a
recoiler, which together form a dipole. The energy and momentum of this
dipole is preserved when a parton, previously on mass shell, is assigned a
virtuality. Preliminary kinematics are constructed directly after each
branching, such that unevolved partons are always on mass shell.

In the final-state shower evolution, the selection of recoilers is based on
colour partners (with colours traced in the $N_C \rightarrow \infty$
limit). A gluon, which carries both colour and anti-colour indices,
therefore has two recoiling partners, with the emission rate split between them.
There are then two further possibilities for selecting the recoiling parton
\begin{Itemize}
\item Initial state: the recoiler is the parton in the initial state that
carries the same\\ (anti-)colour index as the radiator. This dipole
configuration implies that the extra momentum required to keep all partons
on shell is taken from the beam and as a result will be slightly suppressed
by PDF factors. 
\item Final state: the recoiler is the parton in the final state that
carries the opposite\\ (anti-)colour as the radiator (i.e. if the radiator
carries colour index $i$, its recoiler is the parton that carries
anti-colour $i$).
\end{Itemize}
Once a dipole system has been selected to radiate, the kinematics of the
new outgoing partons $b$ and $c$ are constructed in the rest frame of the
radiator ($a$) and recoiler ($r$) system. With the new virtuality given by
$Q^2 = m_a^2 = \pTs/z(1-z)$ and $m_{ar}^2 = (p_a + p_r)^2$, the increase in
$E_a$ and therefore the decrease in $E_r$ is given by
\begin{equation}
\frac{m_{ar}}{2} \rightarrow \frac{(m_{ar}^2 \pm Q^2)}{2m_{ar}} ~.
\end{equation}
Finally, partons $b$, $c$ and the new recoiler are boosted back to their
original frame.

With the initial-state shower, we are now doing a backwards evolution,
where a parton $b$ is resolved as coming from a previous branching $a
\rightarrow bc$. At any branching scale, there will be two resolved
incoming partons to a given interaction, and it is the parton on the
other side of the event that is marked as the recoiler. Parton $b$,
previously massless, is now assigned a spacelike virtuality $m_b^2 =
-Q^2$ while the dipole mass $m_{br}$ remains unchanged
by the branching. Once the kinematics of the branching has been
constructed, there is an extra step to be taken; all partons from the
original $b$ + $r$ interaction must be boosted and rotated so that it is
now parton $a$ that is incoming along the $z$ axis.

Note that this choice is different than in the Catani-Seymour dipole 
picture \cite{Catani:1996vz}. There, the colour flow is used to assign
a recoil parton that can be either in the initial state, as above, or 
in the final state. The drawback of this approach, which at first glance
appears attractive and symmetric with the FSR procedure, can be 
illustrated with the production process $\q\qbar \to \Z^0$. In a first
emission $\q \to \q \g$, indeed the $\Z^0$ takes the $\pT$ recoil of
the emitted gluon. Thereafter, however, the colour flows from the 
beams to the gluon, and the $\Z^0$ is not affected by any subsequent
emissions. This runs counter to the result of a Feynman-diagrammatic
analysis of how the  $\pT$ of the $\Z^0$ can build up by repeated
emissions from the incoming partons. In this case we prefer to 
stay closer to the latter picture, although this may also have its own
shortcomings.

\subsection{Beam Remnants, Primordial $\kT$ and Colour Reconnection}
\label{sec:BR}
When the $\pT$ evolution has come to an end, the beam remnants will
consist of the remaining valence content of the incoming hadrons as well as
any companion quarks. These remnants must carry the remaining fraction of
longitudinal momentum. \tsc{Pythia} will pick $x$ values for each component
of the beam remnants according to distributions such that the valence
content is ``harder'' and will carry away more momentum. In the rare case
that there is no remaining quark content in a beam, a gluon is assigned to
take all the remaining momentum.

The event is then modified to add primordial $\kT$. Partons are expected to
have a non-zero $\kT$ value just from Fermi motion within the incoming
hadrons. A rough estimate based on the size of the proton gives a value of
$\sim 0.3 \GeV$, but when comparing to data, for instance the $\pT$
distribution of $\Z^0$ at CDF, a value of $\sim 2 \GeV$ appears to
be needed. The current solution is to decide a \kT value for each
initiator parton taken from a hadron based on a Gaussian whose width is
generated according to an interpolation
\begin{equation}
\sigma(Q, \widehat{m}) = 
\frac{Q_{\frac{1}{2}} \, \sigma_{\mrm{soft}} + Q  \, \sigma_{\mrm{hard}}}%
{Q_{\frac{1}{2}} + Q } \, 
\frac{\widehat{m}}{\widehat{m}_{\frac{1}{2}} + \widehat{m}}  ~,
\end{equation}
where $Q$ is the hardness of a sub-collision ($\pT$ for a $2 \to 2$
QCD process) and $\widehat{m}$ its invariant mass, $\sigma_{\mrm{soft}}$ 
and $\sigma_{\mrm{hard}}$ is a minimal and maximal value, and 
$Q_{\frac{1}{2}}$ and $\widehat{m}_{\frac{1}{2}}$ the respective scale 
giving a value halfway between the two extremes. Beam remnants are assigned
a separate width $\sigma_{\mrm{remn}}$ comparable with $\sigma_{\mrm{soft}}$.
The independent random selection of primordial $\kT$ values gives a
net imbalance within each incoming beam, which is shared between 
all initiator and remnant partons, with a reduction factor 
$\widehat{m} / ( \widehat{m}_{\frac{1}{2}} + \widehat{m} )$ for initiators of 
low-mass systems. With the  $\kT$'s of the two initiators of a system
known, all the outgoing partons of the system can be rotated and  
Lorentz boosted to the relevant frame. During this process, the
invariant mass and rapidity of all systems is maintained by appropriately
scaling the lightcone momenta of all initiator partons.

The final step is colour reconnection. In the old MPI framework,
good agreement to CDF data is obtained if 90\% of additional
interactions produces two gluons with ``nearest neighbour'' colour
connections \cite{Field:2002vt}. More recently, an annealing algorithm
has been used \cite{Sandhoff:2005jh,*Skands:2007zg}, again requiring
a significant amount of reconnection to describe data.
In \tsc{Pythia 8} colour reconnection is currently performed by giving 
each system a probability to reconnect with a harder system
\begin{equation}
\mathcal{P} = \frac{{\pT}_{Rec}^2}{({\pT}_{Rec}^2 + \pT^2)}, ~~~~~
~~~~~ {\pT}_{Rec} = R \times \pTo,
\label{eqn:crec}
\end{equation}
where $R$ is a user-tunable parameter and $\pTo$ is the same
parameter as in eq.~(\ref{eqn:pt0}). 

The idea of colour reconnection can be motivated by noting that MPI 
leads to many colour strings that will overlap in physical space, which 
makes the separate identity of these strings questionable. Alternatively, 
moving from the limit of  $N_C \rightarrow \infty$ to $N_C = 3$, it is 
not unreasonable to allow these strings to be connected differently 
due to a coincidence of colour. Adapting either of these approaches, 
dynamics is likely to favour reconnections that reduce the
total string length and thereby the potential energy. 

With the above probability for reconnection, it is 
easier to reconnect low-$\pT$ systems, which can be viewed as them having 
a larger spatial extent, such that they are more likely to overlap with 
other colour strings. Currently, however, this is only a convenient 
ansatz. More than that, given the lack of a firm theoretical basis, 
the need for colour reconnection has only been established within the 
context of specific models.

\subsection{Tuning and Energy Dependence}

\textsc{Pythia} contains a large number of parameters, and it is not 
feasible to attempt to tune them all. Even with the less relevant ones 
retained at their default values, the task of tuning the rest simultaneously
is nontrivial. A simplification is offered by the assumption of jet 
universality, since then $\e^+ \e^-$ data can be used to tune FSR and 
hadronisation independently of other aspects. Such a tune exists, closely 
related to the Professor/Rivet tuning of \textsc{Pythia 6.4} 
\cite{Buckley:2009bj}.
With these parameters fixed, a second step is to tune the other aspects,
such as ISR and MPI, to hadron collider data. Here, the \textsc{Pythia 8} 
models differ enough from the previous versions, that carrying across 
parameters is non trivial. Until recently, default parameters for these 
parts of the simulation were based on an early and primitive comparison 
with data. It is only recently that a somewhat more detailed tune, 
``Tune 1'', has been released, based on the Perugia tunes \cite{Skands:2009zm}.
Studies have also been made with the Professor/Rivet framework 
\cite{Hoeth:2009zz}, but have not yet resulted in a complete tune.

For MPI, and therefore also for rescattering, the
$E_{\mrm{CM}}^{\mrm{pow}}$ parameter of eq.~(\ref{eqn:pT0scaling}) is vital
for extrapolating to LHC energies. Prior to Tune 1, a default value of
$E_{\mrm{CM}}^{\mrm{pow}} = 0.16$ was chosen, supported by toy model
studies \cite{Dischler:2000pk} and previous tuning efforts
\cite{Buttar:2005}. The newer Perugia \cite{Skands:2009zm} and Professor
tunes \cite{Buckley:2009bj} now point towards a larger value of
$E_{\mrm{CM}}^{\mrm{pow}} = 0.24$ (as did the earlier Tunes A and DW
\cite{Field:2005sa,Field:2006iy,Field:2009zz}, but based on fewer data),
which has been adopted as the new default value in Tune 1. Such a change
leads to large differences in the extrapolation to LHC energies.

A second difference is the matter profile of the hadron. In the original
model, a double Gaussian distribution was found to best reproduce the data,
where a more ``lumpy'' proton gives rise to more fluctuations from the MPI
framework. In \cite{Skands:2009zm}, it is noted that the introduction of
showers from MPI scattering systems gives rise to a new source of
fluctuations such that it is possible to describe the same data with a
single Gaussian matter distribution. It should be noted that this extra
shower activity also ties in with the increase in
$E_{\mrm{CM}}^{\mrm{pow}}$ above.

The differences between these tunes represents an uncertainty in our
predictions. To give some indication of this uncertainty in what follows,
certain data will be presented using both of these tunes, where, for
simplicity, they will be referred to as the ``old'' and ``new'' tune
respectively. Of course, these parameters are not changed in isolation and
other parameters in the spacelike showers, MPI and beam remnants handling
have been shifted accordingly. More details on the current status of
\textsc{Pythia 8} tunes are available in the documentation contained within
the package.

\section{Rescattering}
\label{sec:rescattering}

An event with a rescattering occurs when an outgoing state from one
scattering is allowed to become the incoming state of another scattering.
In the simplest case, one incoming parton to a scattering will be taken
from the beam, while the other will come from a previous interaction
(a single rescattering). There is, however, the possibility that both
incoming partons will be already scattered partons (a double rescattering).
If we accept MPI as real, then we should also allow rescatterings
to take place. They would show up in the collective effects of MPI,
manifesting themselves as changes to multiplicity, $\pT$ and other
distributions. Unfortunately, these effects may not be so easy to detect in
real life; some of the effects of rescattering will already be accounted
for by the tuning of the existing models, e.g. slight parameter changes in
ISR, FSR and the existing MPI framework. After implementing a model for
rescattering, a retuning of ${\pT}_0$ and other parameters will be
necessary so that rescattering contributions are not double counted. After
such a retuning, it is likely that the impact of rescattering will be
significantly reduced, so we should therefore ask whether there are more
direct ways in which rescattering may show up. Is there perhaps a region of
low $\pT$ jets, where an event is not dominated by ISR/FSR, where this
extra source of three-jet topologies will be visible? A further
consideration is that such rescatterings will generate more $\pT$ in the
perturbative region, which may overall mean it is possible to reduce the
amount of primordial $\kT$ and colour reconnections necessary to match
data, as discussed in Sec.~\ref{sec:BR}.

A comparison between double parton scattering and a single rescattering
process is given in an analytical calculation by Paver and Treleani
\cite{Paver:1983hi,*Paver:1984ux}. Specifically, they look at processes of
the type illustrated in Fig.~\ref{fig:rescattering-schematic}, which are
both of the same order in $\alphas$, but differ in the number of
partons taken from the incoming hadrons. We will not cover this calculation
in detail, but we make note of one interesting aspect. In the single
rescattering case, the parton which goes on to rescatter is now a
propagator, which introduces a $(1 / p^2)$ term in the scattering
amplitude. As usual, the approximation of taking the residue of the pole
at $p^2 = 0$ can be made, but there is perhaps a slight contradiction
when considering the space-time picture of two Lorentz contracted hadrons
passing through each other. If the virtuality of the propagator gives some
indication of the spatial separation of the two interactions, then one may
expect some kind of characteristic scale given by the hadronic dimension,
which is not taken into account in this approximation.
Nevertheless, it is this same scheme which we will use as a starting
point in our modeling of rescattering that follows. In this way, we can
assume that a probabilistic description of rescattering can be used, as is
already done for the existing MPI framework.

Some results of these analytical calculations are given in
Sec.~\ref{sec:results}, but here we point out that this is a fully
parton level calculation and does not include any radiative or
hadronisation effects. The goal within the complete generation framework is
not only to include all such effects, but also to model ``higher-order''
rescatterings in the same way that many MPI processes are included beyond
DPS.

\subsection{Rescattering and PDFs}
\label{sec:rescatteringPy8}
To understand the basic principles of rescattering, we begin with the
typical case of small-angle $2 \to 2$ scattering mediated by a 
$t$-channel gluon exchange. An incoming parton extracted from the 
hadron beam will still exist in the final state, only slightly disturbed 
in momentum space, and can still be associated with the same hadron beam.
In the limit where the angle goes to zero, this parton has the same $x$
value as the incoming one, previously selected according to the appropriate 
PDF. This parton thus contributes to the overall PDF of the beam with
a delta function at the selected $x$ (and with the given flavour), so that the
new PDF can be written as
\begin{equation}
f(x,Q^2) \rightarrow
f_{\mrm{rescaled}}(x,Q^2) + \delta(x - x_1)
\end{equation}
where $f_{\mrm{rescaled}}$ is the original PDF of the hadron, now rescaled to
take into account the flavour and momentum taken from it, and $x_1$ is the 
$x$ value of the extracted parton. In such a picture, the momentum sum 
should still be conserved
\begin{equation}
 \int_0^1 x \left[ f_{\mrm{rescaled}}(x, Q^2) + \delta(x - x_1) \right]
 \d x = 1  ~.
 \label{eqn:sumrule}
\end{equation}
It can be viewed as a quantum mechanical measurement of the wave function
of the incoming hadron, where the original squared wave function 
$f(x,Q^2)$ in part collapses by the measurement process of one of the 
partons in the hadron; that is, one degree of freedom has now
been fixed, while the remaining ones are still undetermined. All the
partons of this disturbed hadron can scatter, and so there is the
possibility for the already extracted parton to scatter again.

We can generalise the above reasoning in two ways: first, to extract more 
than just one parton from the hadron and second, to move beyond only small
angle gluon exchange, such that overall we may write the modified PDF as
\begin{equation}
f(x,Q^2) \rightarrow
f_{\mrm{rescaled}}(x,Q^2) + \sum_{i} \delta(x - x_i)
= f_{\u}(x,Q^2) + f_{\delta}(x,Q^2)
~.
\end{equation}
The sum over delta functions now runs over all partons that are available
to rescatter, including outgoing states from hard/MPI processes and partons
from ISR/FSR branchings, and the subscripts $\u$/$\delta$ refer to the
continuous ``unscattered'' and scattered components respectively. 

When the full range of MPI processes and ISR/FSR branchings are allowed, it
is no longer possible to associate each outgoing parton with a particular
hadron remnant. This means that an appropriate prescription is required to
associate partons with the beam remnants. This is addressed further below.

With the PDF written in this way, the original MPI probability
given in eqs.~(\ref{eqn:pTevol})~and~(\ref{eqn:MPIevol}) can now be
generalised to include the effects of rescattering
\begin{equation}
\frac{\d \mathcal{P}_{\mrm{MPI}}}{\d \pT} \rightarrow
\frac{\d \mathcal{P}_{\mrm{uu}}}{\d \pT} +
\frac{\d \mathcal{P}_{\mrm{u \delta}}}{\d \pT} +
\frac{\d \mathcal{P}_{\mrm{\delta u}}}{\d \pT} +
\frac{\d \mathcal{P}_{\mrm{\delta \delta}}}{\d \pT}
,
\end{equation}
where the uu component now represents the original MPI probability, the
$\mrm{u \delta}$ and $\mrm{\delta u}$ components a single rescattering and
the $\mrm{\delta \delta}$ component a double rescattering. In this way,
rescattering interactions are included in the common $\pT$ evolution of
MPI, ISR and FSR. 

This interleaving introduces a certain amount of coherence. For instance, 
it is possible for an outgoing parton from one interaction to branch, 
with one of the daughters rescattering, but such a branching must occur 
at a scale larger than that of the rescattering. There would not be time
for a shower first to develop down to low scales, and thereafter let one
of those daughter partons rescatter at a high scale.  

We should remind that, as before, the $\pT$ ordering should not be viewed 
as a time ordering but rather as a resolution ordering. What this means 
is that, if viewed in a time-ordered sense, a parton could scatter at a 
high $\pT$ scale and rescatter at a lower one, or the other way around, 
with comparable probabilities. As will become apparent later on, the 
kinematics of scattering, rescattering and showers combined can become 
quite complex, however. Therefore we make one simplification in this 
article, in that we choose to handle kinematics as if the rescattering 
occurs both at a lower $\pT$ \textit{and} a later time than the 
``original'' scattering. The rescattering rate is not affected 
by this kinematics simplification.

\begin{figure}
\begin{center}
\includegraphics[scale=0.5]{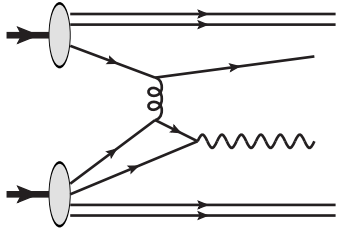}
\end{center}
\caption{$\Z^0$ production with a preceeding rescattering, which is 
not considered in our approach
\label{fig:Z0prod}
}
\end{figure}

This choice should not be a serious restriction 
for the study of jet and UE/MB physics, as is the main objective of this 
article. It does make a difference e.g. for $\Z^0$ production combined 
with a rescattering. Assuming that the $\Z^0$ vertex is at the largest 
scale and therefore defines the original scattering, it would not be 
allowed to have a rescattering that precedes the $\Z^0$ production, 
i.e.\ the ordering illustrated in Fig.~\ref{fig:Z0prod} would be excluded.
Thus, for now, there is  no natural way to study whether the 
rescattering mechanism could be used as a way to reduce primordial $\kT$ 
(Sec.~\ref{sec:BR}).

\subsection{Beam Association}
\label{sec:beamassociation}
\begin{figure}
\begin{center}
\includegraphics[angle=270, scale=0.50]{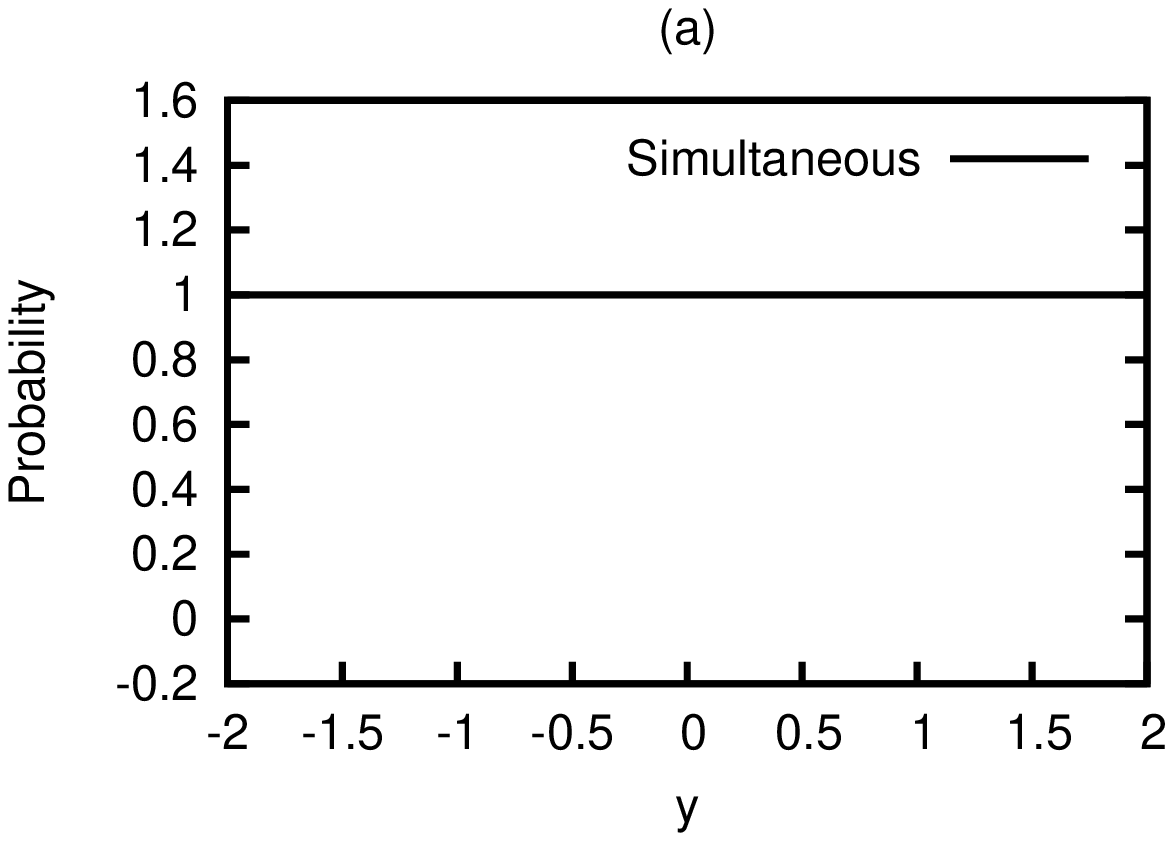}
\includegraphics[angle=270, scale=0.50]{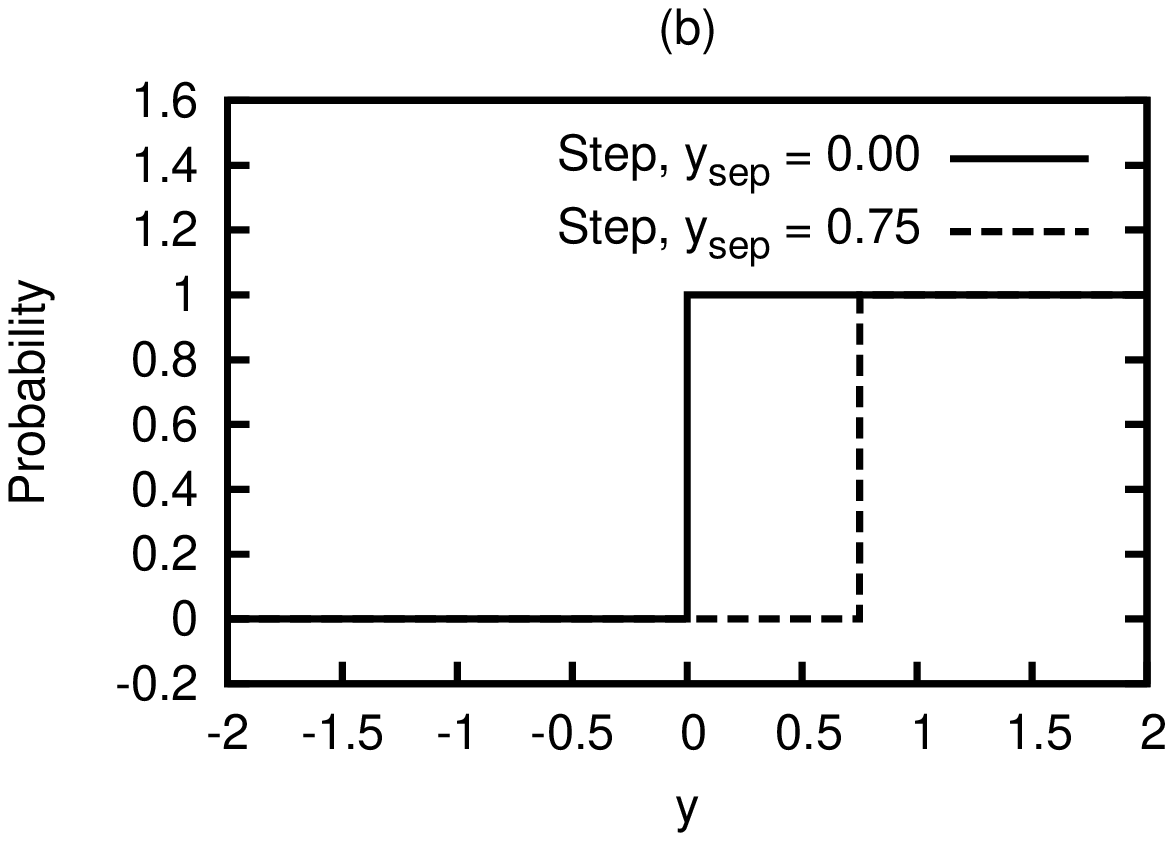}
\includegraphics[angle=270, scale=0.50]{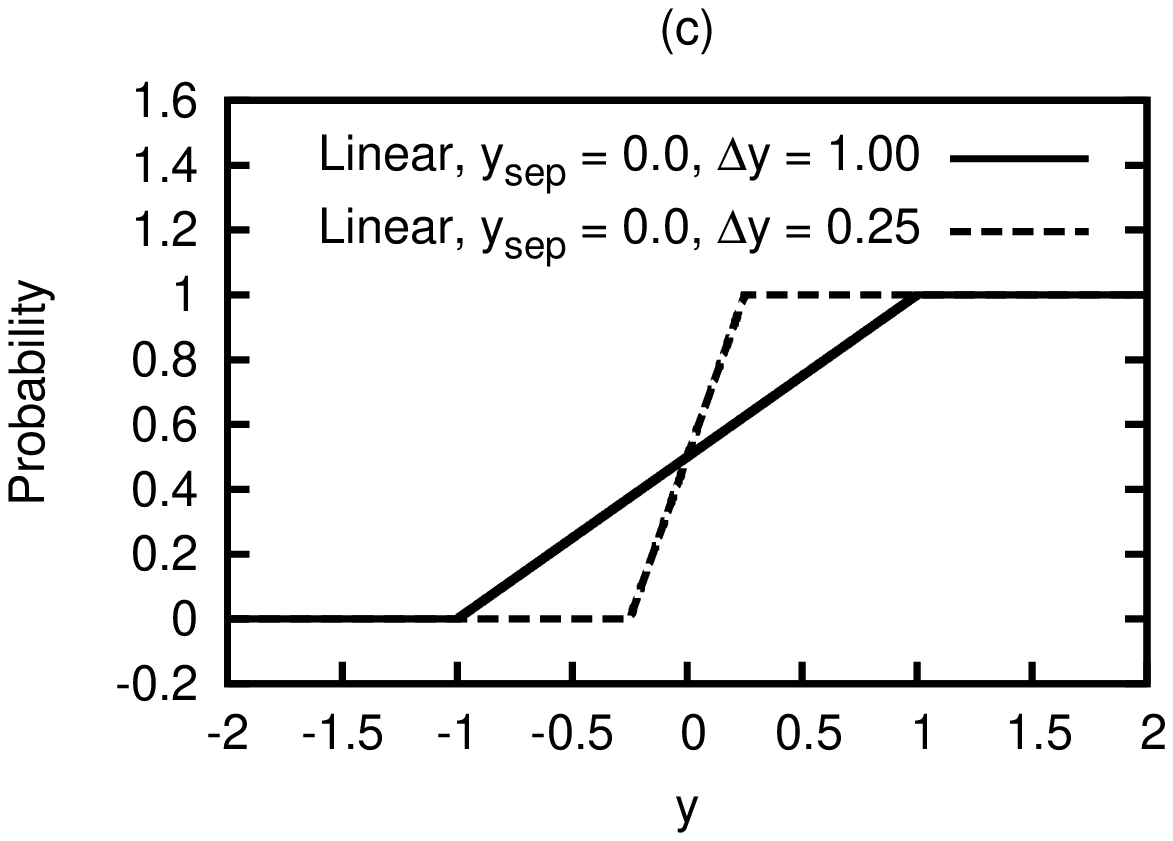}
\includegraphics[angle=270, scale=0.50]{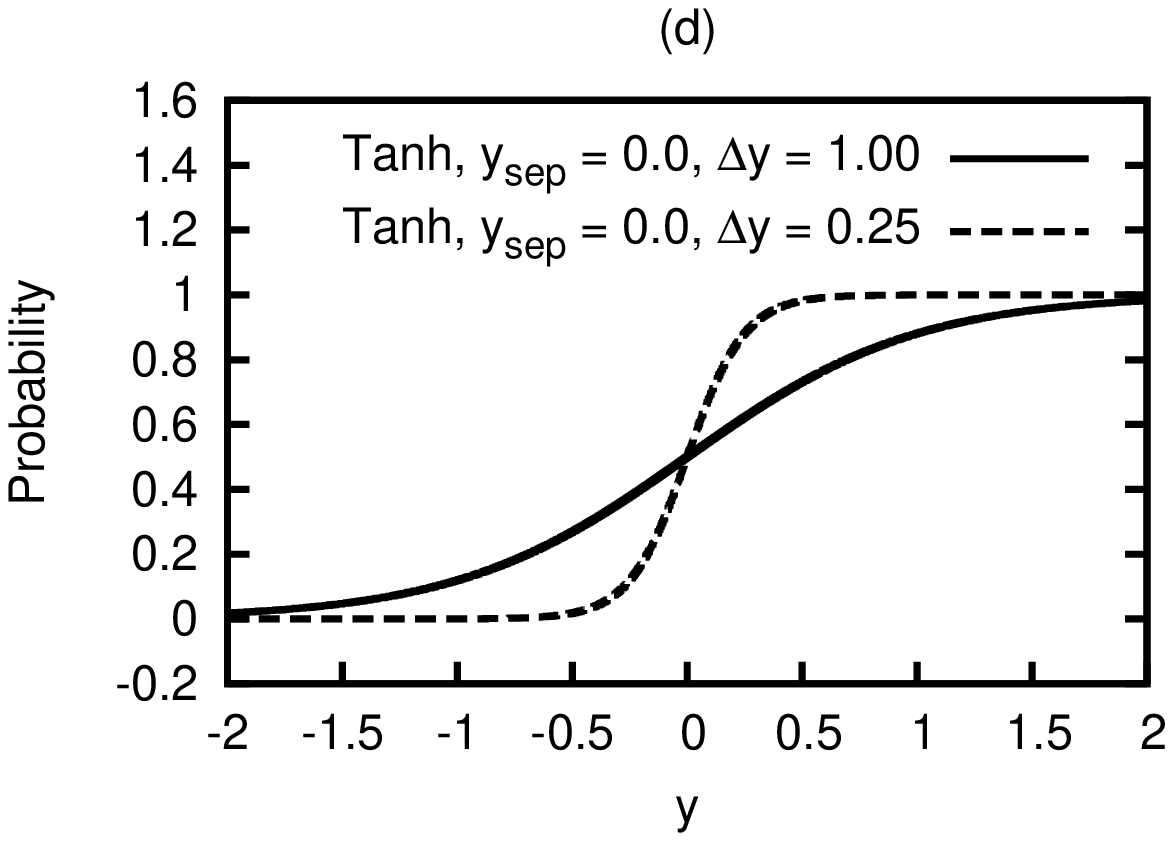}
\end{center}
\caption{Probability of a parton being assigned to beam A for (a)
Simultaneous (b) Step (c) Linear and (d) Tanh
\label{fig:ba-plots}
}
\end{figure}

We now return to the issue of associating scattered partons, that 
potentially may rescatter, with a beam remnant. A parton associated 
with beam A is allowed to rescatter with any of the partons from beam B, 
and vice versa. There are no first principles involved, except that the
description should be symmetric with respect to beams A and B. 
We therefore consider four separate rapidity based prescriptions, some with 
tunable parameters $y_{\mrm{sep}}$ and $\Delta_y$. Expressed in the rest 
frame of the collision, with beam A (B) moving in the $+z$ ($-z$) 
direction, the probability for a parton to be assigned to beam A is
\begin{enumerate}
\item Simultaneous: each parton is treated as belonging to both
incoming beams simultaneously
$$P_A = 1 ~.$$
\item Step: the probability for assignment to beam A is given by
a step function in rapidity with an optional central exclusion or overlap
region specified by $y_{\mrm{sep}}$
$$P_A = \Theta(y - y_{\mrm{sep}}) ~.$$
\item Linear: the probability for assignment to beam A is zero
below $(y_{\mrm{sep}} - \Delta_y)$, unity above $(y_{\mrm{sep}} + \Delta_y)$ 
and rises linearly in between
$$P_A = \frac{1}{2} \left( 1 + \frac{y - y_{\mrm{sep}}}{\Delta_y} \right) ~ ,
~~~~ y_{\mrm{sep}} - \Delta_y < y < y_{\mrm{sep}} + \Delta_y ~ .$$
\item Tanh: the probability for assignment to beam A rises as
$$P_A = \frac{1}{2} \left( 1 + \tanh \left( 
\frac{y - y_{\mrm{sep}}}{\Delta_y} \right) \right) ~.$$
\end{enumerate}
These four scenarios are illustrated in Fig.~\ref{fig:ba-plots}, which
shows the probability of a parton being assigned to beam A for various
parameter settings. 

At first glance, it may appear that the simultaneous
option will lead to many more rescatterings, as essentially twice the
number of partons will be made available to rescatter compared to e.g.
the step function with $y_{\mrm{sep}} = 0$. Specifically, consider the 
single rescattering cross section for a given rescattering parton
\begin{equation}
\frac{\d\sigma}{\d\pTs} = \sum_{i} \int \d x_1 \,
f_i(x_1, Q^2) \, \frac{\d \hat{\sigma}}{\d\pTs} ~.
\label{eq:dsigmadptsingle}
\end{equation}
For $t$-channel gluon exchange, $\d \hat{\sigma} / \d\pTs$ is almost
independent of the mass of the colliding system. This requires, however,
that the squared invariant mass fulfills $\hat{s} \geq \pT^2 / 4$,
else obviously the cross section vanishes. The consequence is that the
allowed $x$ integration range will be smaller when the rescattering 
parton is in the same hemisphere as the yet to be selected parton,
and larger when they are in opposite ones. The difference is quite 
marked in the small-$x$ region, where the PDFs peak, resulting in only
a modest increase in the single rescattering rate for the simultaneous
option.

For double rescattering with the simultaneous option, a doubling of the 
number of interactions is immediately expected, relative to the step 
function, just from interchanging the partons assigned to beam A and 
beam B. A further (approximate) doubling of the number of pairs comes
from considering also those with the two incoming partons in the
same hemisphere. At first glance, one might guess that these latter ones
again would be suppressed by the mass requirement. However, note that
this time there is no integration down to small $x$ involved, since
the incoming parton momenta are already fixed. Furthermore, since the 
evolution is done in order of decreasing $\pT$, the incoming partons were
both produced at a larger $\pT$ scale than is now considered for the 
rescattering. Therefore, already the separation in the transverse momentum 
plane is typically enough to fulfill the $\hat{s}$ condition, even without 
the further separation of longitudinal momenta.

Finally, the linear and tanh options can be seen as smeared-out versions 
of the step one, where one could argue the tanh option as being more 
physical since it contains no discontinuities. 

\begin{figure}
\begin{center}
\includegraphics[angle=270, scale=0.60]{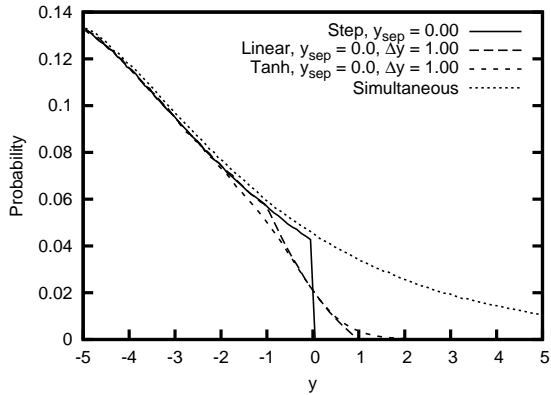}
\end{center}
\caption{Probability of a parton being assigned to Beam B and being
rescattered in LHC minimum bias events ($\p\p$, $\sqrt{s}=14 \TeV$, 
old tune)
\label{fig:LHC-mb-probBeamB}
}
\end{figure}

\begin{figure}
\begin{center}
\includegraphics[angle=270, scale=0.60]{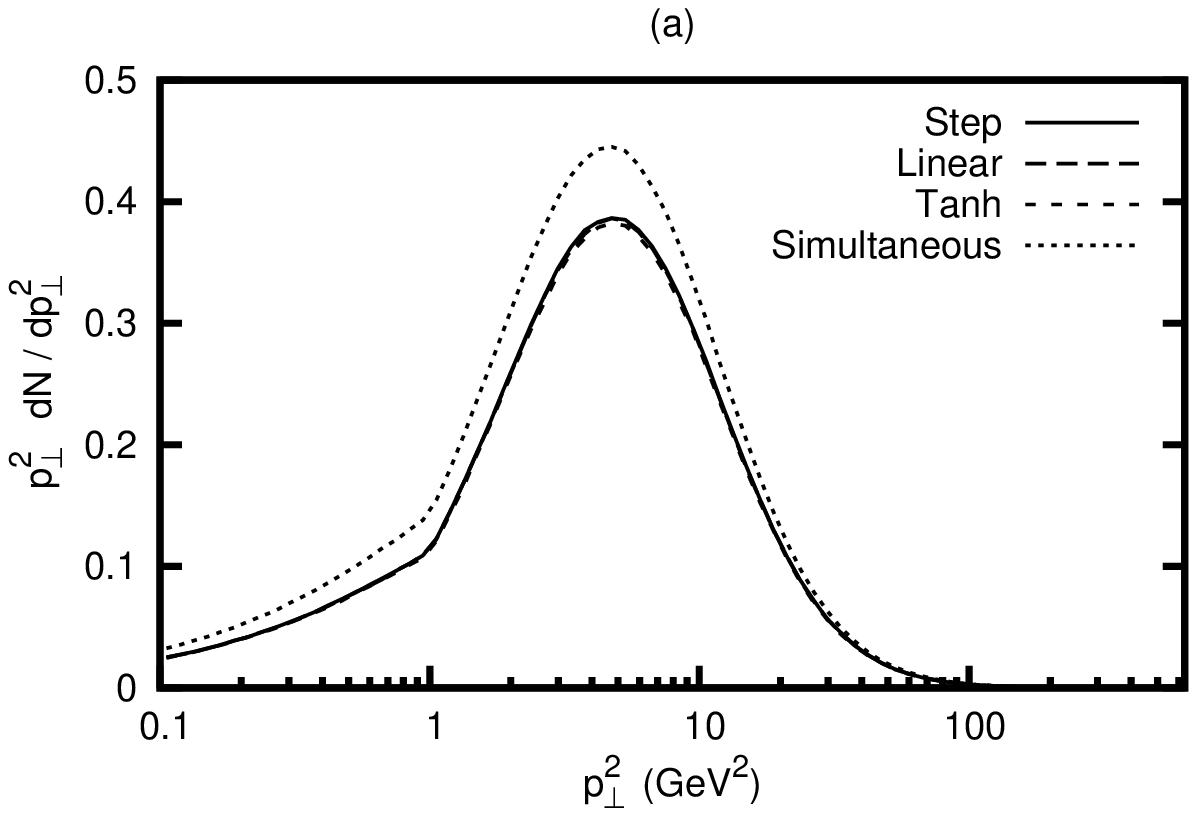}
\includegraphics[angle=270, scale=0.60]{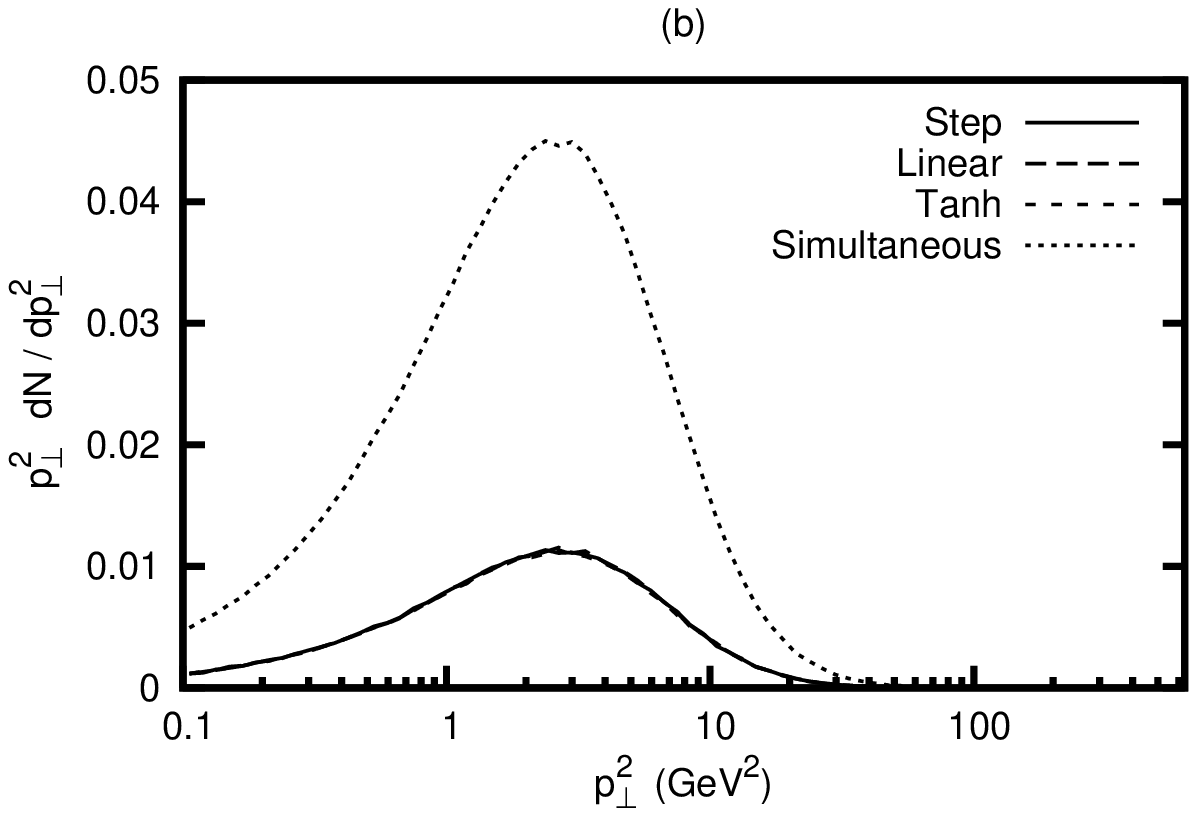}
\end{center}
\caption{$\pT$ distributions of (a) single rescatterings and (b) double
rescatterings in LHC minimum bias events ($\p\p$, $\sqrt{s}=14 \TeV$,
old tune). Parameters for the different options are the same as the
previous figure. Note the difference in vertical scale between (a) and (b),
and that the step, linear and tanh curves are so close that they may be
difficult to distinguish
\label{fig:LHC-mb-pT}
}
\end{figure}

To get an idea of the effects of the various options, we study some
features of LHC minimum-bias events ($\p\p$, $\sqrt{s} = 14.0 \TeV$, old
tune) when no showers are present. Fig.~\ref{fig:LHC-mb-probBeamB}, shows
the probability for a parton to be assigned to beam B and subsequently to
be rescattered as a function of rapidity. The natural suppression in the
simultaneous scenario, as discussed above, is visible in the positive
rapidity region.  Fig.~\ref{fig:LHC-mb-pT} now shows the $\pT$
distributions of (a) single rescatterings and (b) double rescatterings. With
the simultaneous option the increase in single rescatterings is small, as
expected. It is the factor four rise in double rescattering that now stands
out; as explained above there is almost no suppression in interactions where 
the two partons sit in the same hemisphere.

\begin{figure}
\begin{minipage}[c]{0.5\linewidth}
\centering 
\includegraphics[angle=270,scale=0.60]{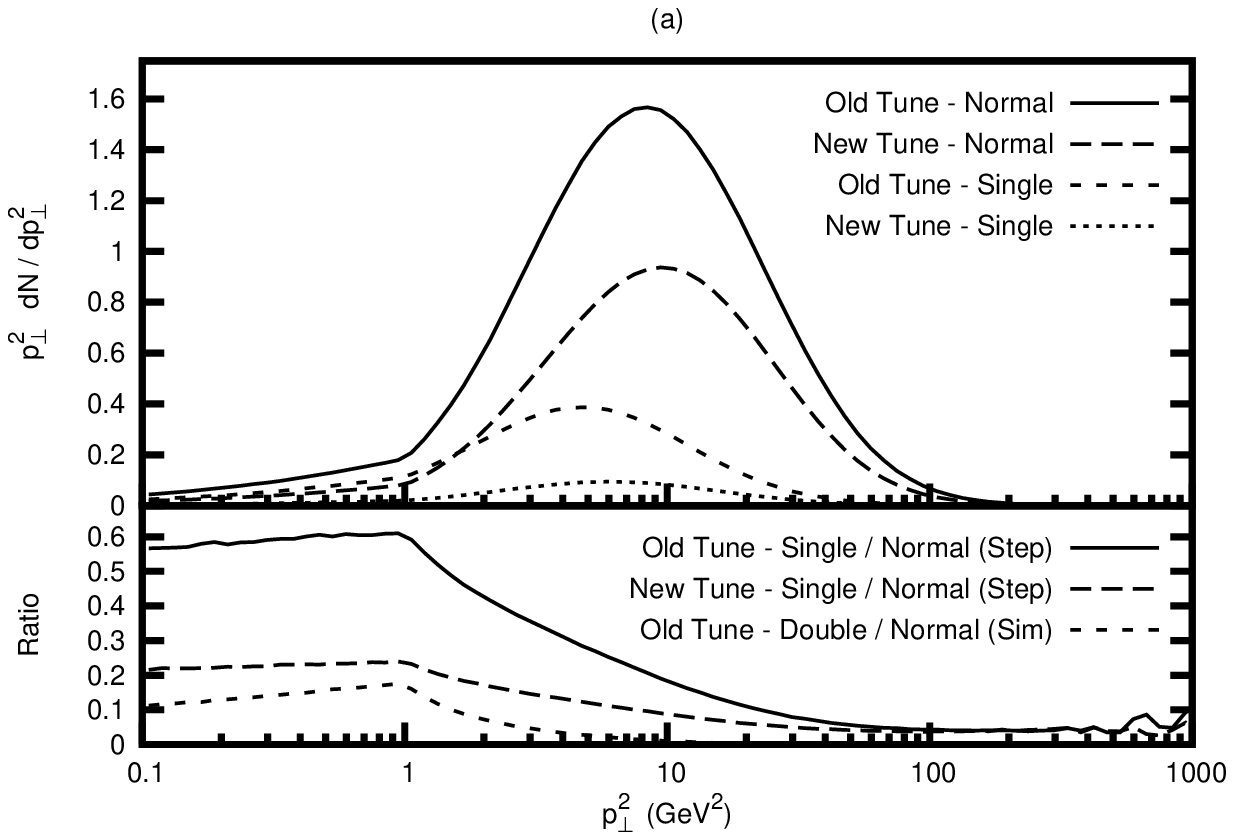}
\end{minipage}
\begin{minipage}[c]{0.5\linewidth}
\centering 
\includegraphics[angle=270,scale=0.60]{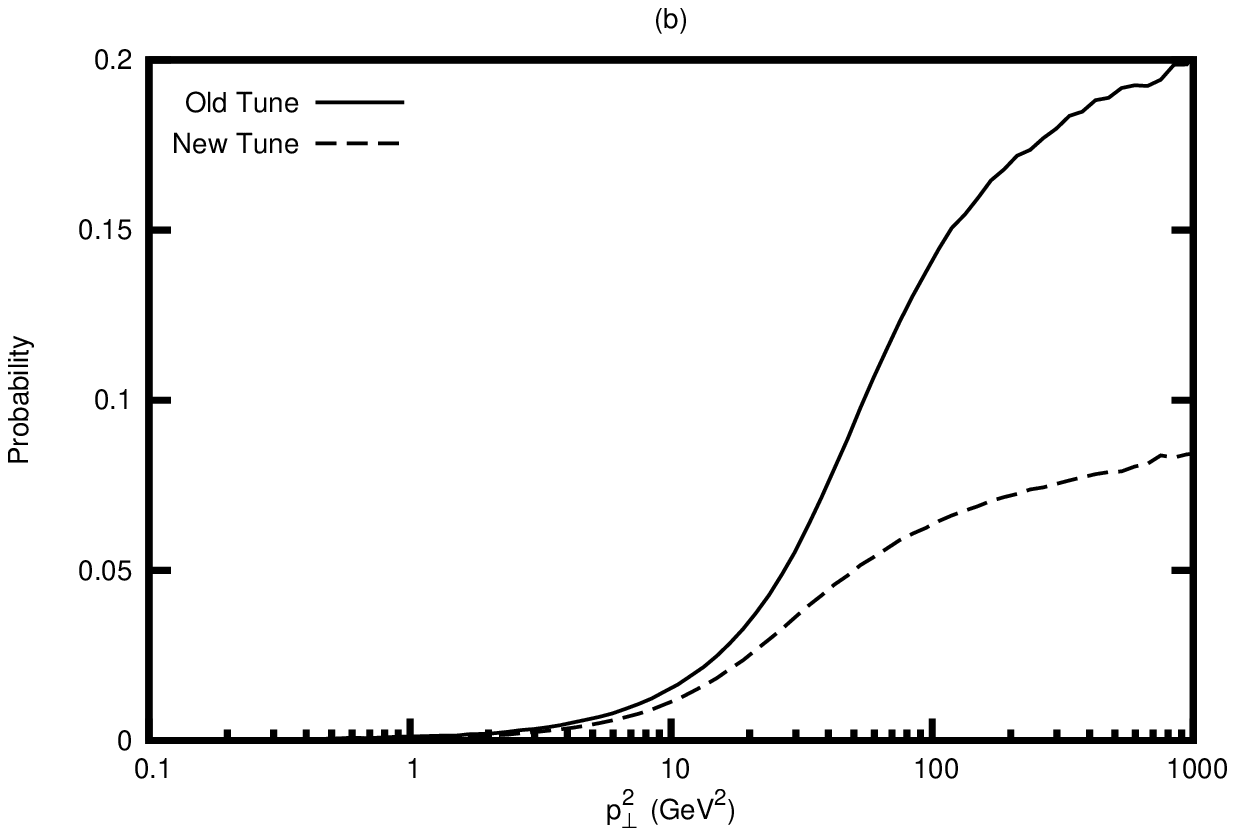}
\end{minipage}
\caption{Rescattering in LHC minimum bias events ($\p\p$, $\sqrt{s} = 14
\TeV$, old and new tunes). (a) shows the $\pT$ distribution of scatterings
and single rescatterings per event. Included in the ratio plot is double
rescattering with the simultaneous beam prescription using the old tune,
where its effect is maximal. (b) shows the probability for a parton,
created in the hard process of an event, to rescatter as a function of its
initial $\pT$
\label{fig:pTres}
}
\end{figure}

At this stage, it is clear that the different beam prescriptions do not
have a large influence on the outcome of rescattering, except for double
rescattering when used with the simultaneous option. In
Fig.~\ref{fig:pTres}a, the $\pT$ distributions of normal MPI scatterings
are shown compared to those of single rescattering for both the old and new
\textsc{Pythia 8} tunes. The effect of the different tunes on the
extrapolation of the MPI model to LHC energies is immediately apparent.
As previously predicted, rescattering is a small
effect at larger $\pT$ scales, but, when evolving downwards, its relative
importance grows as more and more partons are scattered out of the incoming
hadrons and become available to rescatter. The suppression of the cross
section at small $\pT^2$ is caused mainly by the regularisation outlined in
eq.~(\ref{eqn:pt0}), but is also affected by the scaling violation in the
PDFs. Below $\pT^2 \sim 1 \GeV^2$ the PDFs are frozen, giving rise to an
abrupt change in slope. Normal scatterings dominate, but there is a clear
contribution from single rescatterings. Double rescattering is too small to
be visible in the upper plot, but included in the lower plot is
the ratio of double rescattering to normal rescattering for the
simultaneous option, where the old tune has been used to generate the
maximum effect possible. Even in this maximal case, the growth of double
rescattering with lowering $\pT$ is slow, peaking around the 10\% level in
the low-$\pT$ region where it is likely that any effects will be ``washed
out'' by other low $\pT$ activity.

Although, in this formalism, rescattering is a low-$\pT$ effect (insofar as
it occurs at low scales in the $\pT$ evolution of the event), we
point out that it can have an effect on the high-$\pT$ properties of an
event. Fig.~\ref{fig:pTres}b shows the probability for a parton created in
the hard process of an event to go on and rescatter as a function of the
initial $\pT$ of the parton. A parton created at a high $\pT$ will have
a larger range of $\pT$ evolution, meaning that there is a greater chance
that it will rescatter at some point in this evolution.

Finally, as an indicator of the effect of energy on the growth of
rescattering, Table~\ref{tab:noRes} shows the average number of scatterings
and rescatterings for different types of event at Tevatron and LHC energies
(step option only, old and new tunes). With all these points in
mind, from now on we no longer consider double rescattering effects and
restrict ourselves to the step beam prescription; with just these
options, the implementation is simplified while the bulk of the interesting
phase space region is still covered.

\renewcommand{\arraystretch}{1.15}
\begin{table}
\begin{center}
\begin{tabular}{c|l|c|c|c|c|}
\cline{3-6}
\multicolumn{2}{c|}{}
& \multicolumn{2}{|c|}{\textbf{Tevatron}}
& \multicolumn{2}{|c|}{\textbf{LHC}} \\
\cline{3-6}
\multicolumn{2}{c|}{}
& \textbf{Min Bias} & \textbf{QCD Jets} & \textbf{Min Bias}
& \textbf{QCD Jets} \\
\cline{2-6}
\multirow{3}{*}{\textbf{Old}\quad}
& \textbf{Scatterings}          & 2.81  & 5.09  & 5.19  & 12.19\phantom{0} \\
& \textbf{Single rescatterings} & 0.41  & 1.32  & 1.03  & 4.10             \\
& \textbf{Double rescatterings} & 0.01  & 0.04  & 0.03  & 0.15             \\
\cline{2-6}
\multirow{3}{*}{\textbf{New}\quad}
& \textbf{Scatterings}          & 2.50  & 3.79  & 3.40  & 5.68  \\
& \textbf{Single rescatterings} & 0.24  & 0.60  & 0.25  & 0.66  \\
& \textbf{Double rescatterings} & 0.00  & 0.01  & 0.00  & 0.01  \\
\cline{2-6}
\end{tabular}
\end{center}
\caption{Average number of scatterings, single rescatterings and double
rescatterings in minimum bias and QCD jet events at
Tevatron ($\p\pbar$, $\sqrt{s} = 1.96 \TeV$,
QCD jet $\hat{p}_{\perp \mrm{min}} = 20 \GeV$)
and LHC ($\p\p$, $\sqrt{s} = 14.0 \TeV$,
QCD jet $\hat{p}_{\perp \mrm{min}} = 50 \GeV$)
energies for both the old and new tunes
\label{tab:noRes}
}
\end{table}
\renewcommand{\arraystretch}{1}

\subsection{Inclusion of Radiation and Beam Remnants}

The addition of rescattering has non-trivial effects on the colour flow in
events. Without rescattering, in the $N_C \rightarrow \infty$ limit,
all colours are confined within a $2 \to 2$ scattering 
subsystem. With rescattering, you now have the possibility for colour 
to flow from one system to another, and thereby to form radiating dipoles
stretched between two systems. Should we then expect differences in 
radiation, relative to a normal dipole confined inside a subsystem? A
crude qualitative argument is that, for a rescattering dipole, there are
more propagators sitting between radiator and recoiler than normally,
which, in an average sense, corresponds to a larger spatial
separation. As a result, one may expect a suppression of hard radiation,
with the normal full rate only in the soft limit.

\begin{figure}
\begin{center}
\includegraphics[scale=0.5]{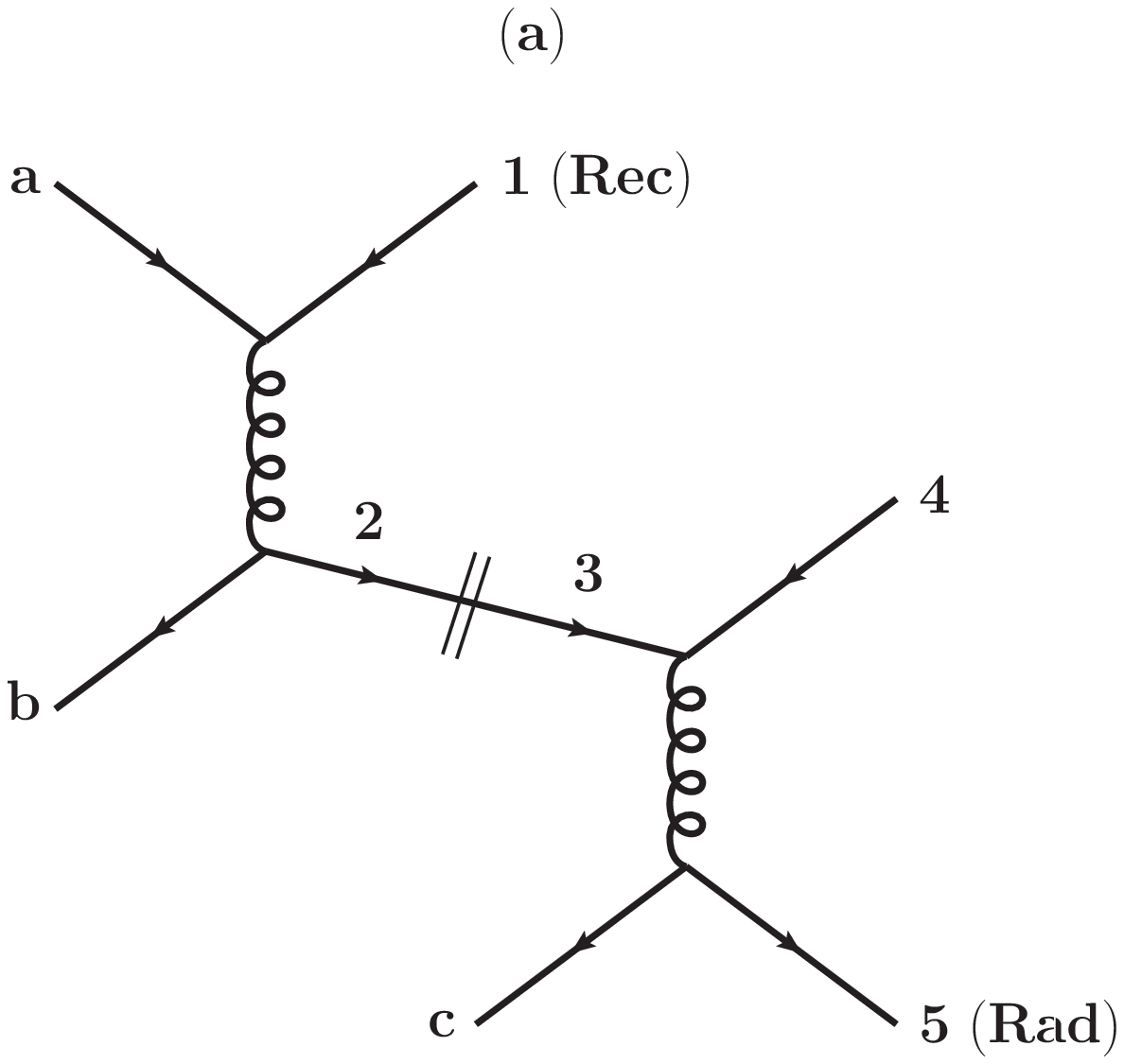}
\hspace{10mm}
\includegraphics[scale=0.5]{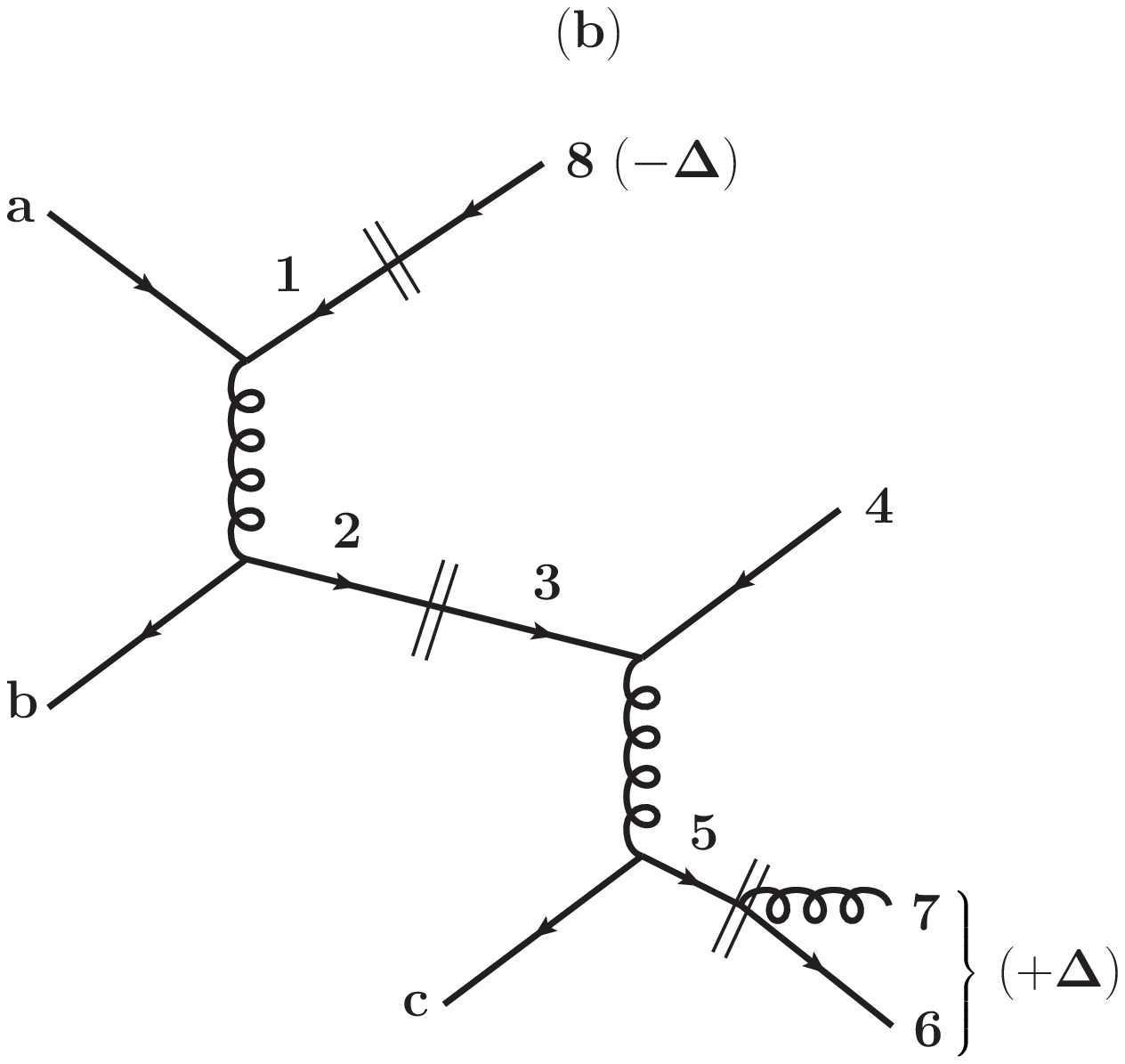}
\end{center}
\caption{Final-state dipole spanning two scattering subsystems (a) before
branching and (b) after branching. $\Delta$ is the four-momentum
transferred from the recoiler to the radiator and the double lines indicate
where a new particle has been bookkept
\label{fig:FSR-dipole}
}
\end{figure}

This is not the only new issue when showers and beam remnants are to be
combined with the rescattering concept. There are considerable technical 
complications when we consider those steps of event generation that rely 
on Lorentz boosts to kinematically shift partons, specifically ISR and 
primordial $\kT$. For instance, consider a situation where a final-state 
dipole spans two systems, such as in Fig.~\ref{fig:FSR-dipole}. In
(a), partons 1 and 2 are outgoing from the first system, parton 3 is the
rescattering parton now incoming to the second system (identical to parton
2, but marked separately for clarity) and partons 4 and 5 are outgoing from
the second system. We now consider what happens when parton 5 radiates a
gluon with parton 1 as its recoiling partner. In the existing framework,
the situation after the branching is shown in (b); parton 5 has branched
into partons 6 and 7, while parton 8 now reflects the changed kinematics of
the recoiler.  After the branching, both systems combined
still conserve energy and momentum, but individually they do not.

Alone, this is not a significant issue. However, the introduction of 
initial-state branchings and primordial $\kT$ means that the incoming 
partons $a$,  $b$ and $c$ time and again receive a changed four-momentum. 
More specifically, it is this $\pT$ kick combined with longitudinal momentum 
and energy transfer that guarantees that the invariant mass of the 
$a+b$ and $c+3$ systems are preserved. However, while 
$p_c + p_3 = p_4 + p_5$, such that any $\pT$ kick on $c$ is transferred
to the $4+5$ final state, after the dipole emission 
$p_c + p_3 \neq p_4 + p_6 + p_7$. Thereby the kick on the final state will
become wrong, leading to an overall imbalance in an event. The effect is 
especially big if the original lightcone momentum of $c$ is very small, 
while that of 1, along the same axis, is large. Then a modest transfer 
from 1 to $6+7$ can give a big multiplication factor for a $\pT$ kick on $c$.  

\begin{figure}
\begin{center}
\includegraphics[scale=0.5]{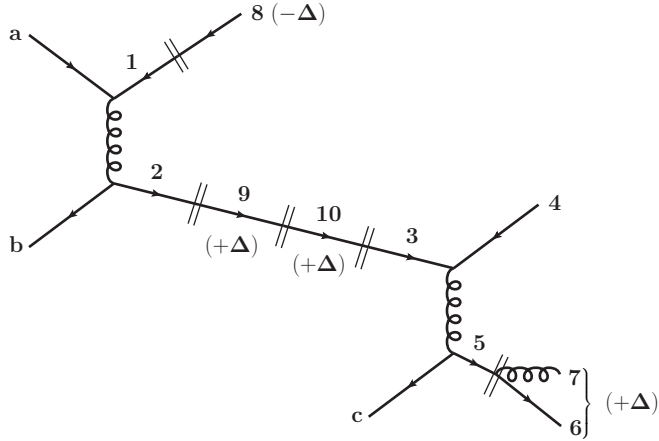}
\end{center}
\caption{Branching of a final-state dipole spanning two scattering subsystems
where the internal line connecting the two systems has been
changed to conserve momentum
\label{fig:FSR-dipole2}
}
\end{figure}

The solution we have adopted is, for FSR, to propagate
momentum shifts through all intermediate systems, such that energy and
momentum are conserved locally in all systems. Fig.~\ref{fig:FSR-dipole2}
shows how the branching of Fig.~\ref{fig:FSR-dipole} now looks with this
scheme. It is now partons 8 and 9 that are outgoing from the first system,
while it is parton 10 and no longer parton 3 that is incoming to the second
system. 

The example shown is one of the simplest possible; there can be one or more 
intermediate systems sitting between the
radiating and recoiling systems, such that it is more than one intermediate
line that must have its momentum changed. It is also important to note that
the momentum transfer has already been defined by the procedure of keeping
the dipole mass unchanged. This means that, when propagating the recoil
along all internal lines connecting the systems, the partons are going to 
acquire nonzero virtualities, spacelike or timelike.

This scheme does allow for the complete parton shower and hadronisation
framework to be used, but does have disadvantages. Firstly, this procedure
is only valid when a path can be found that connects the radiating and
recoiling systems, which for QCD radiation will almost always be
true. When there is no direct path between the two different systems, the
emission is vetoed. In principle this procedure may then be carried
out for all branchings that meet the above criteria, but in practice
we identify the following cases where additional vetoes are required:
\begin{Itemize}
\item Negative system $\hat{s}$: this scheme can result in both the
invariant masses and rapidities of systems being altered. While we
explicitly allow this to happen, there is the restriction that the
$\hat{s}$ of a system may not turn negative. Any branching that would cause
such a situation is therefore vetoed.
\item Too large virtuality or negative energy: Lorentz transformations
mix the energy and momentum components of a four-vector. Allowing the
virtuality of an incoming or outgoing parton to become large in
comparison to the invariant mass of its system, or allowing an outgoing
parton to have an arbitrarily large negative energy would result in
partons gaining a large ``unphysical'' $\pT$ kick from a boost. This
kick is compensated, so that momentum is conserved, but the transverse
energy of the event may increase more than it should.
Unfortunately, there are no simple criteria for deciding when a boost
becomes too large such that it is deemed improper. The limits used
are therefore to veto emissions when they cause a parton virtuality or
energy (in the rest frame) to come within a certain fraction of the system
invariant mass.
\end{Itemize}

\begin{figure}
\begin{center}
\includegraphics[angle=270,scale=0.65]{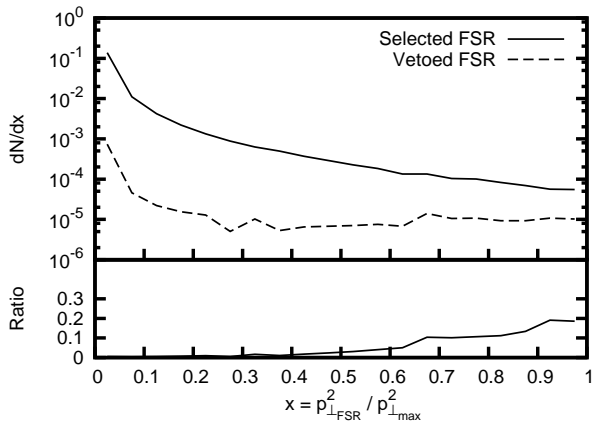}
\end{center}
\caption{Amount of FSR vetoed by momentum propagation procedure (see text).
LHC jet events ($\p\p$, $\sqrt{s} = 14 \TeV$, $\hat{p}_{\perp \mrm{min}} =
100 \GeV$, $40 \leq {\pT}_{\mrm{max}} \leq 50 \GeV$, old tune)
\label{fig:FSR-Fail}
}
\end{figure}

The suppression of FSR caused by this scheme is shown in
Fig.~\ref{fig:FSR-Fail} for LHC jet events ($\p\p$, $\sqrt{s} = 14.0 \TeV$,
$\hat{p}_{\perp \mrm{min}} = 100 \GeV$, old tune). The upper plot shows the
ratio of $\pT^2$ values for final-state radiation to the maximal $\pT^2$
value (i.e. the creation scale of the system), for all selected $\pT^2$
values and for those that are then subsequently vetoed as above. Only
radiation from rescattering systems produced with $40 \leq \pT \leq 50
\GeV$ is considered for clarity. From the ratio plot, it is evident that
the amount of radiation being vetoed per system is not large, reaching
around 20\% in the high-$\pT$ region, and that the pattern also follows
what may be expected by the reasoning given above for FSR suppression; a
perhaps not-unfortunate side effect of the vetoing procedure. Even with
this suppression in place, there are still effects on those stages of the
simulation that use Lorentz boosts. We return to these after first giving
some details on ISR and primordial $\kT$.

To study ISR, again consider the situation in Fig.~\ref{fig:FSR-dipole}a. 
One case is when one of the incoming partons to a system is from the beam, 
and therefore can radiate within the usual backwards evolution picture, 
while the other is a rescattering parton, such as $c + 3$. As
discussed in Sec.~\ref{sec:MPIinPythia8.ISR.FSR}, these two incoming 
partons define an effective dipole. An initial-state emission will change 
the momentum of $c$. This situation can be handled in much the same way 
as in a normal $2\to 2$ system, with the difference that parton 3 may have 
nonzero $\pT$ and may also have a mass so that the kinematics is more 
complicated \cite{Bengtsson:1986gz}.

Extra work is required when the radiating system has outgoing partons
that go on to rescatter, however. That is, if either $a$ or $b$ obtain
changed momentum by ISR, then also 1 and 2 are changed, and hence the 
incoming 3 of the rescattering system, and finally also the outgoing
partons 4 and 5. There is then a choice to be made: to change the momentum 
of the other incoming parton $c$ or not. In the latter case, with  one 
incoming parton changed and the other not, the internal kinematics of the 
$3 + c \to 4 + 5$ collision cannot be preserved. That is, its invariant
mass and collision $\pT$ are changed, both of which we would rather
preserve. Instead, to keep the system mass unchanged, the 4-momentum of the
incoming parton $c$ must be rescaled, up or down. In the case that the
outgoing partons 4 and 5 go on to rescatter this procedure must be applied
again to the new systems, until all daughter systems have had their
kinematics updated. There is a price to pay in this approach, since any
extra energy must be taken from one of the incoming beams. In the worst
case, the extra energy may exceed what is available in the beam remnant.
When this happens, the original hard process is kept, but the shower and
MPI stages of the simulation are restarted.

To generate full events, the machinery of primordial $\kT$ must also be
added. With the scheme outlined above, the boosts used here do not present
any additional issues compared with the ISR boosts above. Only partons
originating from the beams are given primordial $\kT$, and again, any
$\pT$ kick given to a parton that goes on to rescatter will be propagated
to all daughter systems. There is one further point to address, however. As
explained in Sec.~\ref{sec:BR}, the lightcone momenta of the two
initiator partons to a system are rescaled to keep the invariant mass and
rapidity of the system unchanged. If one of the incoming partons is now a
rescattering parton, then it is natural to require its four-momentum to
remain unchanged. If we do this, there are no longer enough degrees of
freedom to keep both $\hat{s}$ and $y$ of the system unchanged. The default
choice, therefore, is to keep $\hat{s}$ unchanged, by a suitable scaling of
the momentum of the incoming beam particle, but to allow the rapidity of the
system to change. It may, however, be undesirable to allow the rapidities
of systems to change, so there is an additional option to allow the
lightcone momentum of the rescattering parton to change, such that
both $\hat{s}$ and $y$ can be maintained.

\begin{figure}
\begin{center}
\includegraphics[angle=270,scale=0.65]{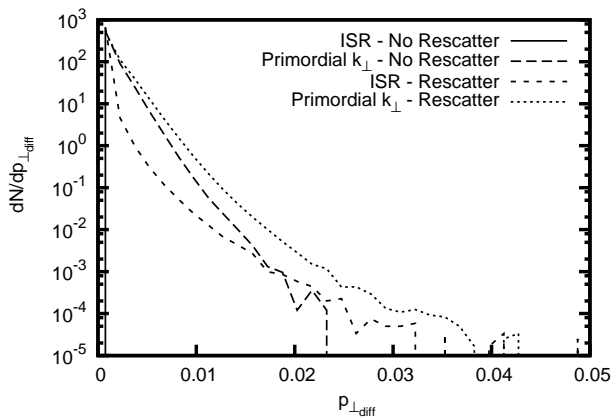}
\end{center}
\caption{Effects of the modified ISR/FSR/MPI scheme on those stages
involving Lorentz boosts. LHC jet events ($\p\p$, $\sqrt{s} = 14 \TeV$, 
$\hat{p}_{\perp \mrm{min}} = 100 \GeV$, old tune).
${\pT}_{\mrm{diff}}$ is defined in eq.~(\ref{eqn:pTdiff})
\label{fig:Boost-Fail}
}
\end{figure}

Finally, we study the potential effects of the above scheme on events.
With rescattering, we expect some deviation from the ``standard'' case for how
Lorentz boosts of ISR and primordial $\kT$ affect events; the question is 
how we can
quantify these deviations. As a metric, for each system undergoing a boost,
we use two quantities, ${\pT}_{\mrm{kick}}$ as a measure of the expected $\pT$
shift of a system and ${\pT}_{\Delta}$ as a measure of the actual $\pT$ shift
of a system, defined as
\begin{eqnarray}
{\pT}_{\mrm{kick}}   &=& (p'_1 + p'_2 - p_1 - p_2)_{\perp} ~,\\
{\pT}_{\Delta} &=& 
\renewcommand{\arraystretch}{0.6} 
\sum_{\begin{matrix} \scriptstyle\mrm{outgoing} \\
\scriptstyle\mrm{after} \end{matrix}} \pT - 
\sum_{\begin{matrix} \scriptstyle\mrm{outgoing} \\
\scriptstyle\mrm{before} \end{matrix}} \pT ~.
\renewcommand{\arraystretch}{1.0} 
\end{eqnarray}
Here, $p_1,p_2$ are the previous incoming four-vectors to a system,
$p'_1,p'_2$ are the new incoming four-vectors to a system and the sums are
over all outgoing partons of a system, after and before the boost has been
performed. Interpreting these measures is not
completely straightforward. For instance, if we consider two back-to-back
jets with transverse momenta only in the $p_y$ direction, then a boost
purely in the $p_x$ direction will result in a ${\pT}_{\Delta} <
{\pT}_{\mrm{kick}}$. Indeed, most of the boosts will have such a configuration,
but a per-event quantity
\begin{equation}
{\pT}_{\mrm{diff}} = 
\frac{ \displaystyle \sum_{\mrm{systems}} 
\max \left( 0, {\pT}_{\Delta} - {\pT}_{\mrm{kick}} \right) }
{\displaystyle\sum_{\mrm{final}} \pT} ~,
\label{eqn:pTdiff}
\end{equation}
where the first sum is over all systems that are boosted in either ISR or
primordial $\kT$, and the second sum over all final state particles, can be used
to give an indication of those events where we may be giving extra unwanted
$\pT$ to certain
systems, which is where the biggest concern lies. The results of this are
shown in Fig.~\ref{fig:Boost-Fail} for LHC QCD jet events ($\p\p$, $\sqrt{s} =
14 \TeV$, $\hat{p}_{\perp \mrm{min}} = 100 \GeV$, old tune), split into
contributions from ISR and primordial $\kT$. An immediate observation is
that, without rescattering, ISR does not give any tail according to this
measure, while primordial $\kT$ does. Here, it is important to note that the
procedure of shifting the frame of a system will often involve rotations
which can skew the $\pT$ of a system and it is the boosts of primordial
$\kT$, where both incoming partons to a system are simultaneously changed,
which are the most extreme. The effects of primordial $\kT$ are 
increased when rescattering is introduced, and also ISR enters the game,
but it is important to note that there is no qualitative change of the
shape. Overall, we find that in around 1 event in 100,000 there is an
extra $\pT$ of around 1\% of the total event $\pT$, which, although not
ideal, is not too worrying.

\section{Results}
\label{sec:results}

\subsection{Parton Level}
\label{sec:resultsParton}
\begin{figure}
\begin{minipage}[c]{0.5\linewidth}
\centering 
\includegraphics[angle=270,scale=0.6]{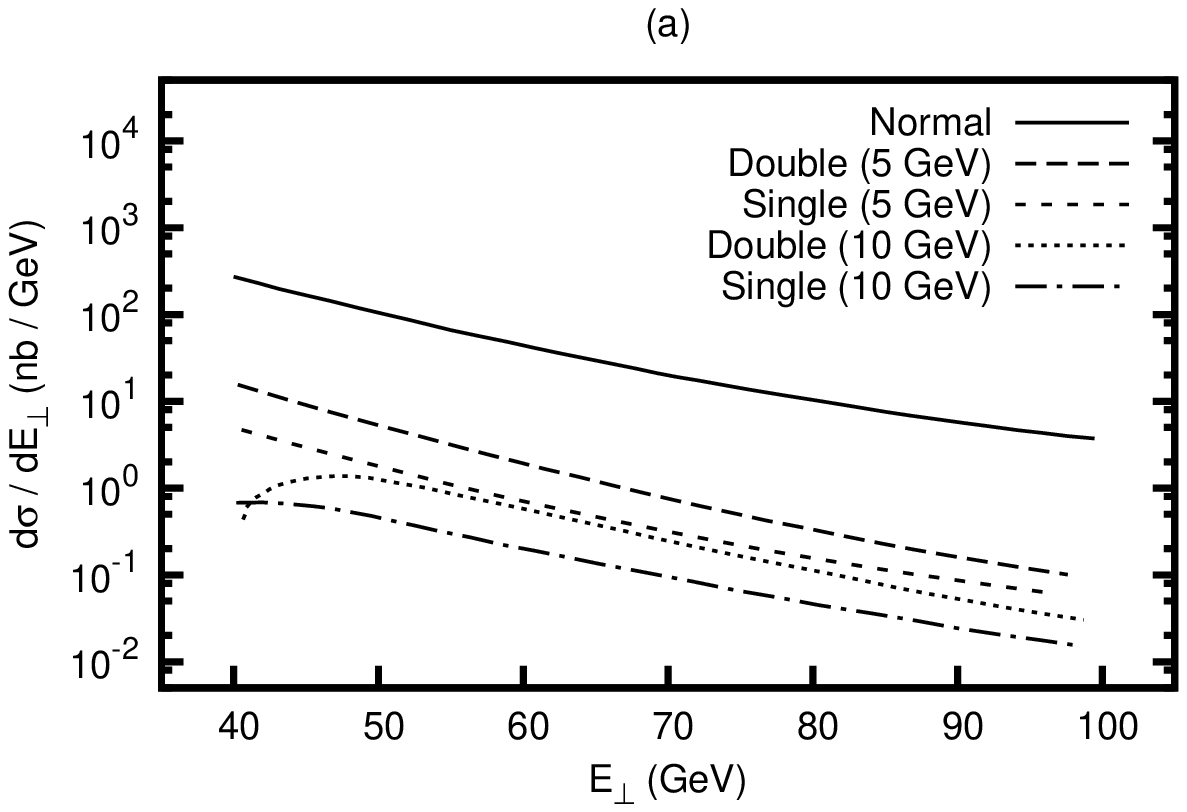}
\end{minipage}
\begin{minipage}[c]{0.5\linewidth}
\centering 
\includegraphics[angle=270,scale=0.6]{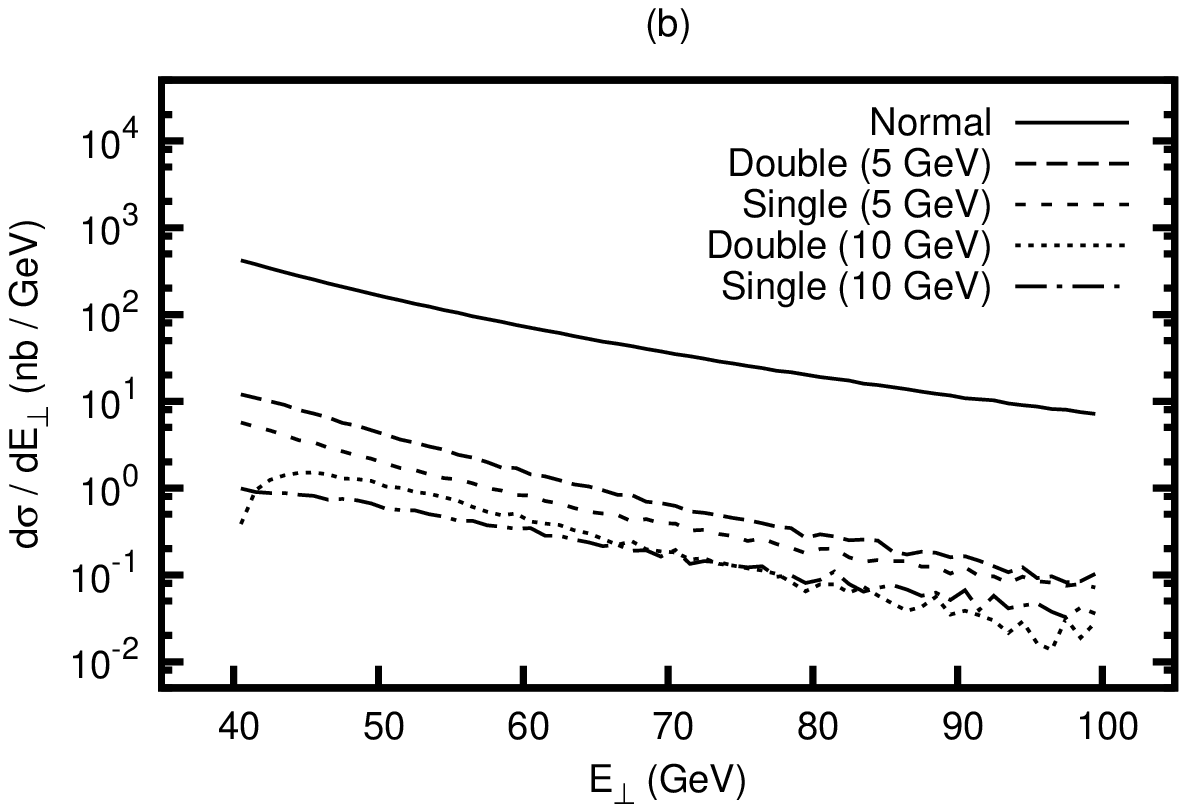}
\end{minipage}
\begin{minipage}[c]{0.5\linewidth}
\centering 
\includegraphics[angle=270,scale=0.6]{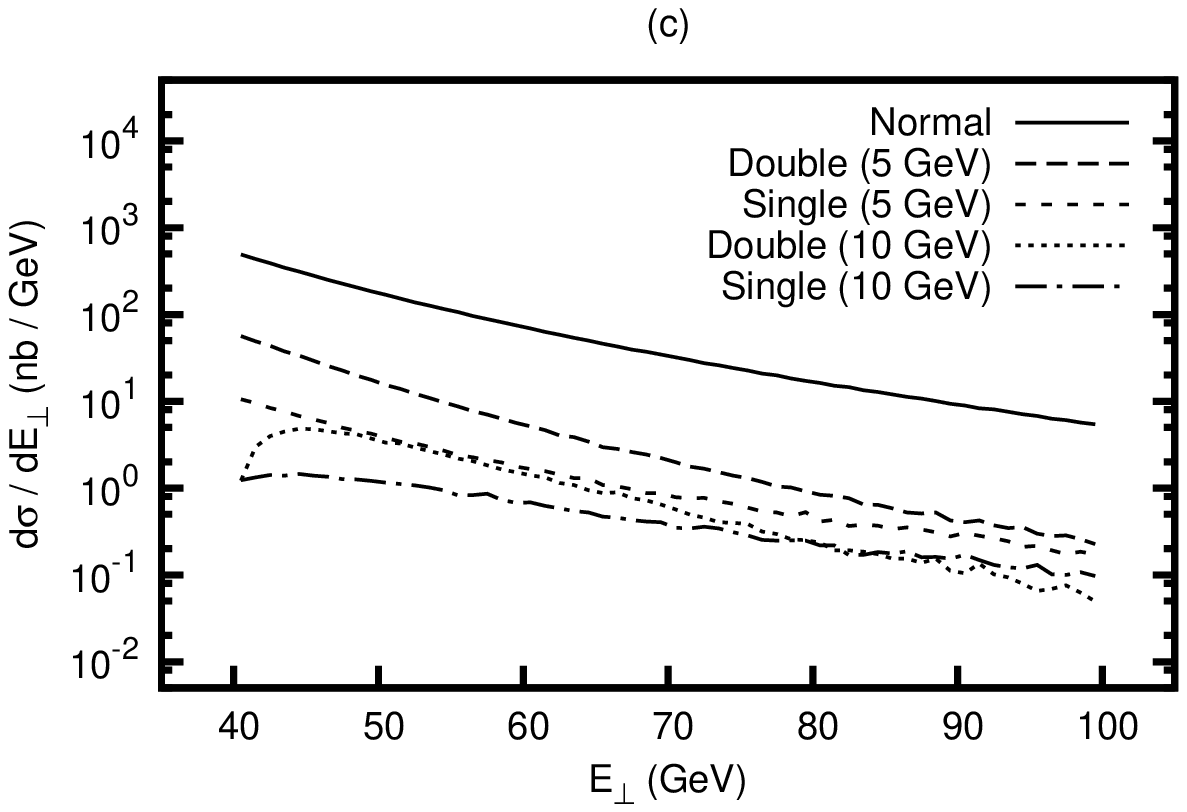}
\end{minipage}
\begin{minipage}[c]{0.5\linewidth}
\centering 
\includegraphics[angle=270,scale=0.6]{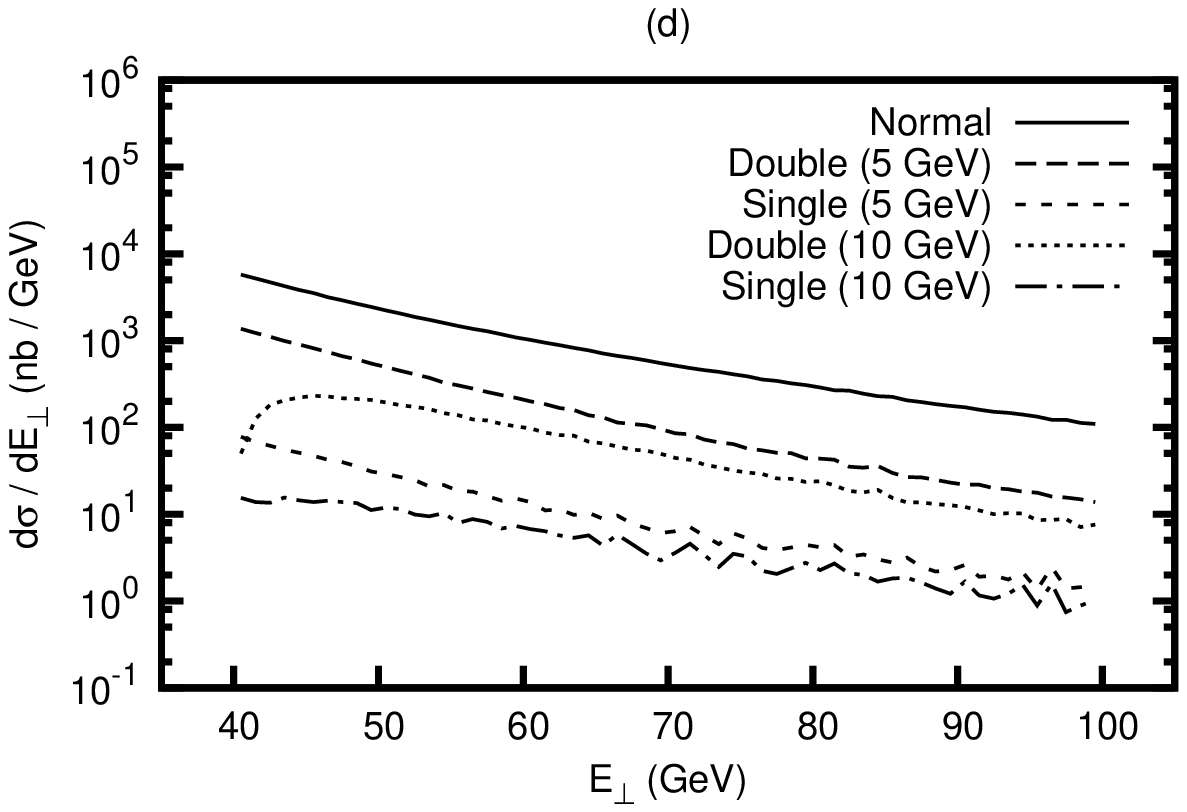}
\end{minipage}
\caption{Parton level jet cross sections with $|y| < 1.0$.
Normal, single and double refer to the two-, three- and four-jet cross
sections respectively. ``$10 \GeV$'' refers to the case where all jets must
have $E_{\perp} > 10 \GeV$ while ``5 \GeV'' refers to the case where one jet
must have $E_{\perp} > 5 \GeV$ and the remaining ones must have
$E_{\perp} > 10 \GeV$. Plots show
(a) analytical results of Paver and Treleani,
$\p\pbar$, $\sqrt{s} = 2.0 \TeV$;
(b) \textsc{Pythia}, $\p\pbar$, $\sqrt{s} = 2.0 \TeV$, 
GHR PDF parameterisation, light QCD only, new tune;
(c) \textsc{Pythia}, $\p\pbar$, $\sqrt{s} = 2.0 \TeV$, CTEQ5L PDF, all heavy
quark effects, new tune;
(d) \textsc{Pythia}, $\p\p$, $\sqrt{s} = 14.0 \TeV$, CTEQ5L PDF, all
heavy quark effects, new tune
\label{fig:Res-Parton}
}
\end{figure}

We begin by returning to the analytical calculation of Paver and Treleani
\cite{Paver:1983hi,*Paver:1984ux}. To get an idea of the relative number of
events for DPS and single rescattering, they integrate their cross sections
from a transverse energy of $40 \GeV$ upwards over a rapidity interval of
$| y | < 1$ for each jet. For both the three-jet and four-jet cross
sections, they look at two cases: one where all jets must have $E_{\perp} >
10 \GeV$ and another where one of the jets is allowed to have $E_{\perp} > 5
\GeV$, while the rest must have $E_{\perp} > 10 \GeV$.  Their results for
Tevatron energies ($\p\pbar$, $\sqrt{s} = 2 \TeV$) are shown in
Fig.~\ref{fig:Res-Parton}a as a function of the summed transverse energy of
the jets, with the two-jet cross section given for comparison. In each
case, the three-jet cross section is smaller than its four-jet counterpart,
but still suggests that rescattering may be a noticeable source of three-jet
topologies.

We now move on to the results from \textsc{Pythia}, where the same cuts
as in Fig.~\ref{fig:Res-Parton}a have been used. We also begin with
Tevatron energies ($\p\pbar$, $\sqrt{s} = 2 \TeV$) using 
the old PDF parameterisation of Gl\"uck, Hoffmann and Reya (GHR)
\cite{Gluck:1980cp} as in the analytical calculation. Both the hard process
and multiparton interaction processes in \textsc{Pythia} are limited to
light QCD ($\u, \d, \s$) only, but the full 1st order running of $\alphas$ has
been left in place, however. Qualitatively, the results from
\textsc{Pythia} are similar to the analytical calculation.

One interesting difference is the larger three-jet cross section (compared
to four jets) when all jets have $E_{\perp} > 10 \GeV$ at high $E_{\perp}$ values.
The key difference between DPS and single rescattering events are the
number of partons taken from the incoming hadrons; four in the former and
three in the latter. Assuming that the hardest $2 \rightarrow 2$ process
has taken place, we are left with the PDF integrals given in
eq.~(\ref{eq:dsigmadpt}) for DPS and eq.~(\ref{eq:dsigmadptsingle}) for
single rescattering. Given the $E_{\perp}$ scales in question, one would
na\"ievly expect the PDF weights for rescattering to be smaller than those
for DPS, but now we must consider the effects of the rapidity cut, $| y | <
1$. With DPS, for both jets to lie within this central rapidity region, the
$x$ integration ranges of the two PDFs are now linked such that a large $x$
value from one PDF factor must be matched by a similarly large $x$ value
from the other. In this way, the four-jet cross section falls off faster
than the three-jet one as we start to consider harder jets in the central
rapidity region.  Taken to the extreme, for MPI jets in a very narrow bin
in rapidity, we would expect the four-jet cross section to be suppressed.
 
We now stay with Tevatron events, but move on to the newer CTEQ5L PDF set,
as well as allowing all possible QCD hard processes and the full set of
default MPI processes.  Heavy quark effects are expected to be small, so
most changes will be driven by differences in the PDFs. A key difference
between the two sets is the gluon content; CTEQ5L has a softer
distribution, making it easier to extract low-$x$ gluons in the
kinematically relevant region. The results are shown in
Fig.~\ref{fig:Res-Parton}c. As expected, both the three- and four-jet cross
sections are higher. The ratio of four- to three-jet cross sections also
has increased slightly, especially at low-$E_{\perp}$, reflecting that the
increased gluon number in both beams makes DPS more favourable.  Finally,
in Fig.~\ref{fig:Res-Parton}d, the results are shown for LHC energies
($\p\p$, $\sqrt{s} = 14 \TeV$, new tune). Both four-jet cross sections now
sit well above their three-jet counterparts and the reasoning as for
Fig.~\ref{fig:Res-Parton}c still holds; the increase in the PDFs of both
beams makes DPS more likely. The results for the old tune are not shown,
but show essentially the same picture; the extra MPI activity leads only to
a slight increase in the three- and four-jet cross sections.

\begin{figure}
\centering 
\includegraphics[scale=0.6]{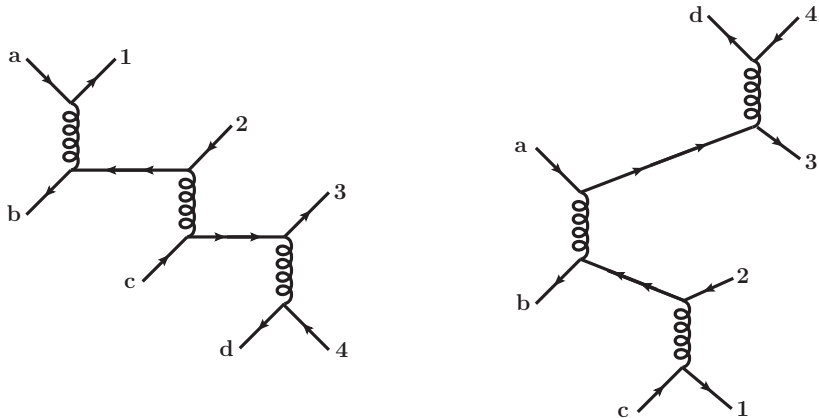}
\caption{Two examples of $4 \rightarrow 4$ topologies which come from two
single rescatterings. The $\pT$ scale of each interaction decreases moving
from left to right and each diagram has four incoming (a, b, c, d) and
four outgoing (1, 2, 3, 4) partons
\label{fig:4to4}
}
\end{figure}

\begin{figure}
\begin{minipage}[c]{0.5\linewidth}
\centering 
\includegraphics[angle=270,scale=0.6]{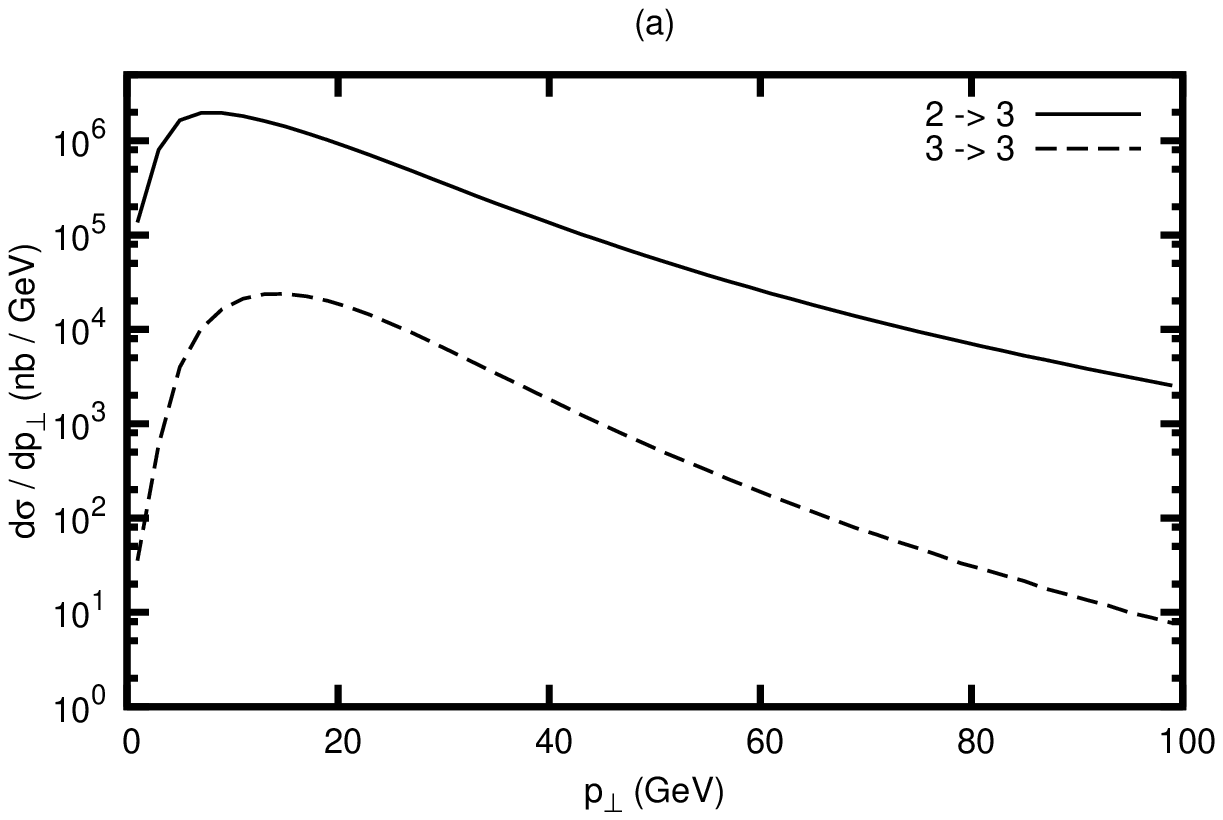}
\end{minipage}
\begin{minipage}[c]{0.5\linewidth}
\centering 
\includegraphics[angle=270,scale=0.6]{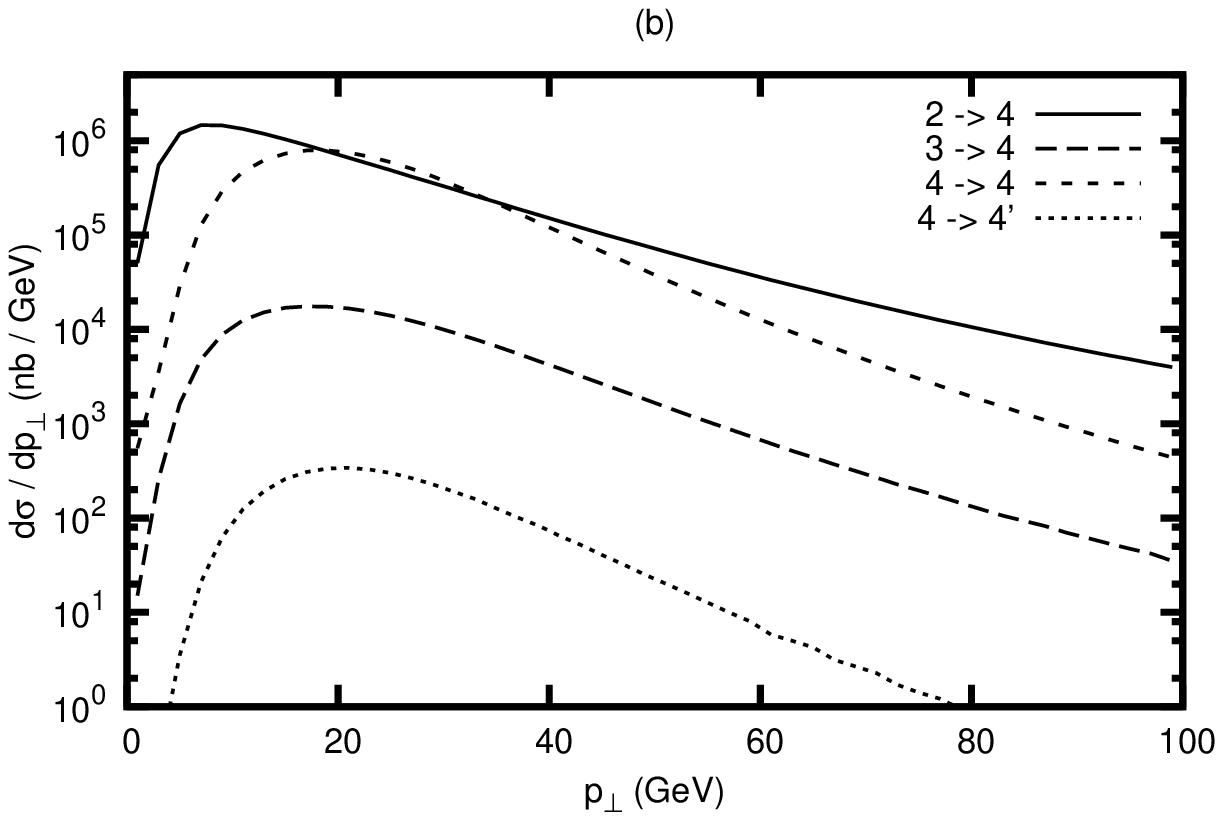}
\end{minipage}
\caption{Breakdown of contributions to the (a) three-jet and (b) four-jet cross
sections (see text) for LHC minimum bias events ($\p\p$, $\sqrt{s} = 14
\TeV$, new tune) when no $\pT$ or rapidity cuts are applied
\label{fig:ResCrossAll}
}
\end{figure}

Unfortunately, this is not the end of the story. Both the three- and four-jet
cross sections have large radiative contributions, and it is to these that
we must compare rescattering. To do this we now run with the full ISR, FSR
and MPI frameworks (but remain at the parton level), and look at the makeup
of the different cross sections. Here we look at events after
the first few steps in the downwards $\pT$ evolution only; although there will
still be contributions to the cross-sections from later stages in the
evolution, we only aim to capture the dominant contributions in the following.
For the three-jet case, we identify the following contributions:
(a) $2 \rightarrow 3$ from single radiation,
(b) $3 \rightarrow 3$ from a single rescattering, and
(c) $4 \rightarrow 3$ from DPS with one jet lost due to cuts.
Instead, for the the four-jet case, we identify:
(a) $2 \rightarrow 4$ from double radiation,
(b) $3 \rightarrow 4$ from single radiation + a single rescattering,
(c) $4 \rightarrow 4$ from DPS, and
(d) $4 \rightarrow 4'$ from two single rescatterings (e.g. Fig.~\ref{fig:4to4}).
The different contributions, as a function of the summed $\pT$ of the jets,
are shown in Fig.~\ref{fig:ResCrossAll} for LHC minimum bias events
($\p\p$, $\sqrt{s} = 14 \TeV$, new tune) when no rapidity or $\pT$ cuts are
applied. Without a rapidity cut, the $4 \rightarrow 3$ configuration is now not
possible. The radiative contributions are dominant over most of the $\pT$
range, but there is a low-$\pT$ region where the DPS cross section is also
large. The $3 \rightarrow 3$ and $3 \rightarrow 4$ rescattering
contributions are suppressed by roughly two orders of magnitude. Finally,
we note that there is a contribution, albeit small, from 2 single
rescatterings leading to a $4 \rightarrow 4'$ configuration.

\begin{figure}
\begin{minipage}[c]{0.5\linewidth}
\centering 
\includegraphics[angle=270,scale=0.6]{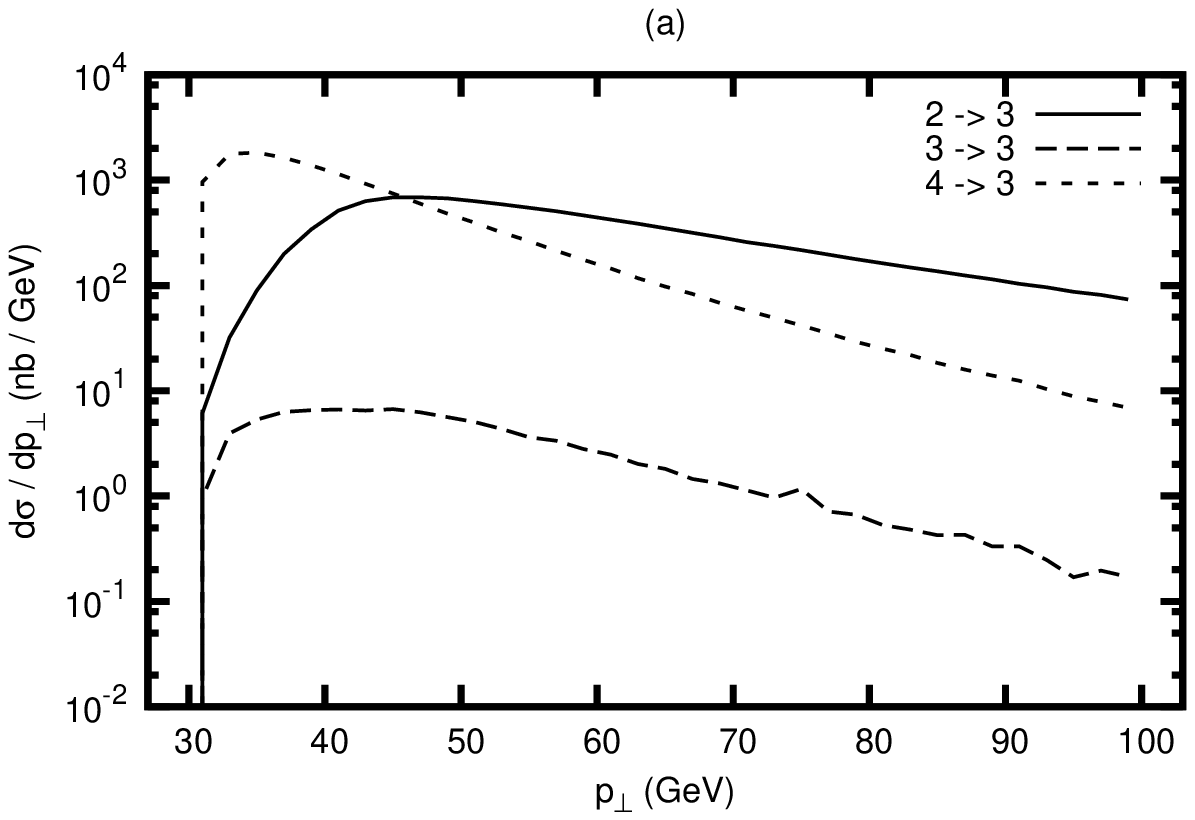}
\end{minipage}
\begin{minipage}[c]{0.5\linewidth}
\centering 
\includegraphics[angle=270,scale=0.6]{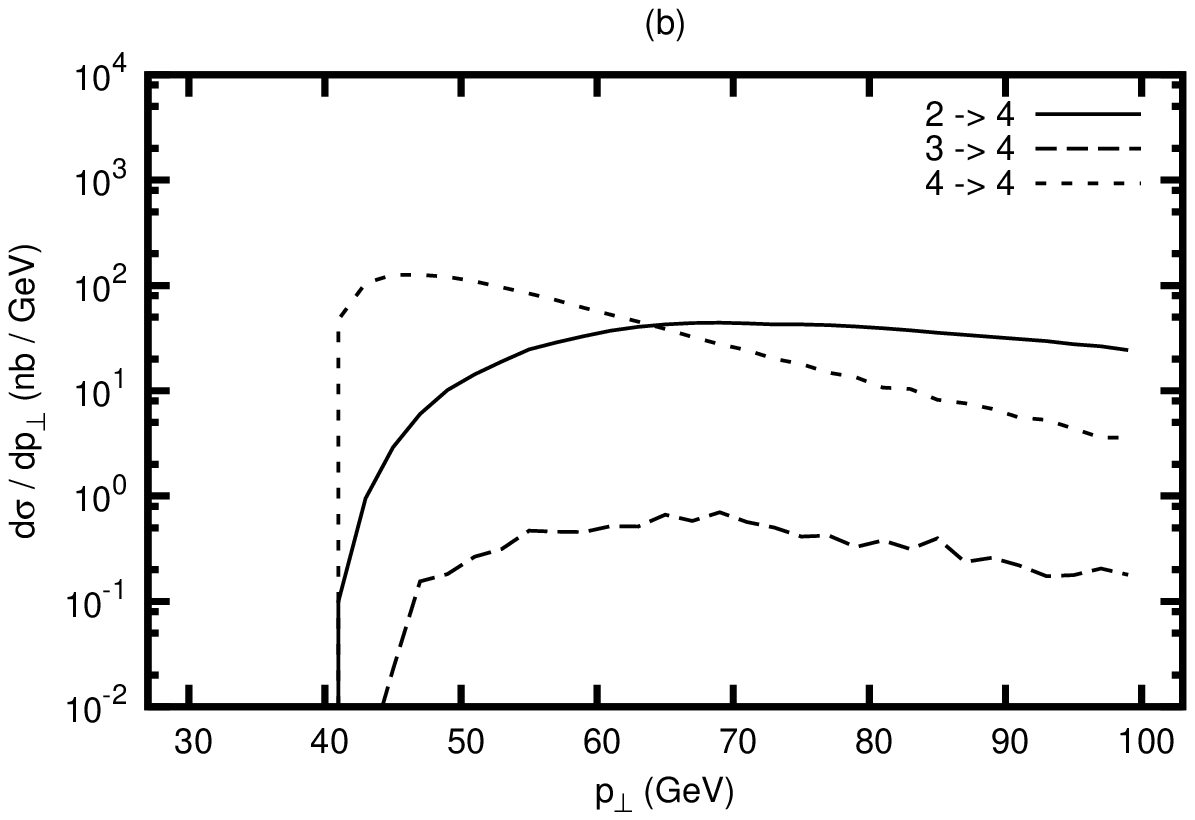}
\end{minipage}
\caption{Breakdown of contributions to the (a) three-jet and (b) four-jet cross
sections (see text) for LHC minimum bias events ($\p\p$, $\sqrt{s} = 14
\TeV$, new tune) with $\pT > 10 \GeV$ and $| \eta | < 1.0$
\label{fig:ResCross}
}
\end{figure}

We now additionally introduce cuts, such that all jets must have a minimum
$\pT$ and lie in some pseudorapidity range. One of the key effects of a
pseudorapidity cut is the addition of a new source of 3-jet events; those
coming from DPS, but where one of the jets does not fall within the allowed 
$\eta$ range. The results for $\pT > 10 \GeV$ and $| \eta | < 1.0 $ are
shown in Fig.~\ref{fig:ResCross}. Immediately it is obvious that those
events where rescattering is present is small compared to the large
background of radiative and DPS events. We also note that the
$4 \rightarrow 4'$ sample is now too small to be visible in
Fig.~\ref{fig:ResCross}b.

The results so far are not too encouraging. The background to single
rescattering is large, but we now move on to hadronic observables; here we
can instead look for signs of the collective effects of the potentially
many rescatterings per event. Further, in Sec.~\ref{sec:jetstudy} we look
to see if there are any kinematical differences which may distinguish
events which contain rescattering.

\subsection{Colour Reconnection}

\begin{figure}
\begin{minipage}[c]{0.5\linewidth}
\centering 
\includegraphics[angle=270,scale=0.60]{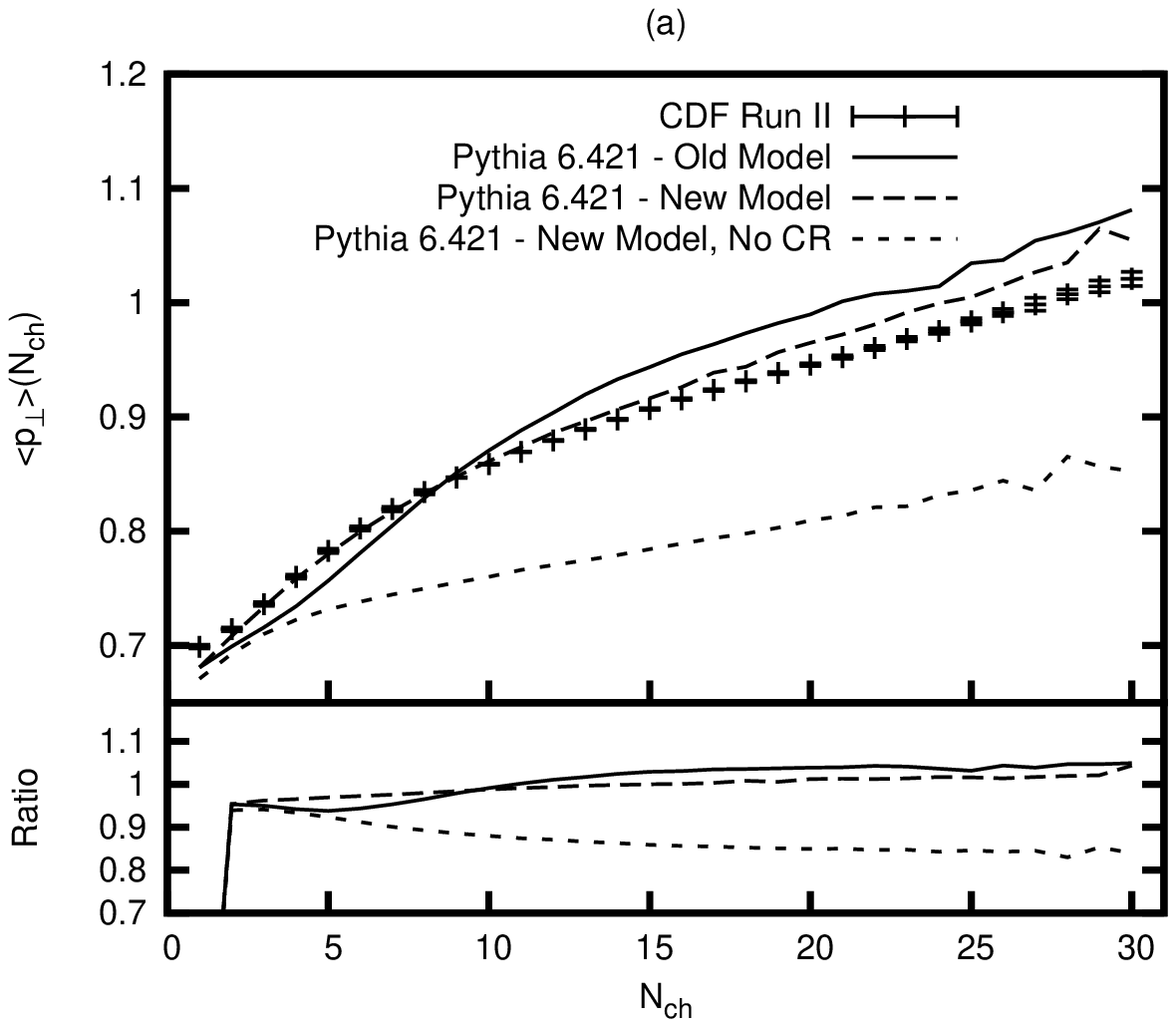}
\end{minipage}
\begin{minipage}[c]{0.5\linewidth}
\centering 
\includegraphics[angle=270,scale=0.60]{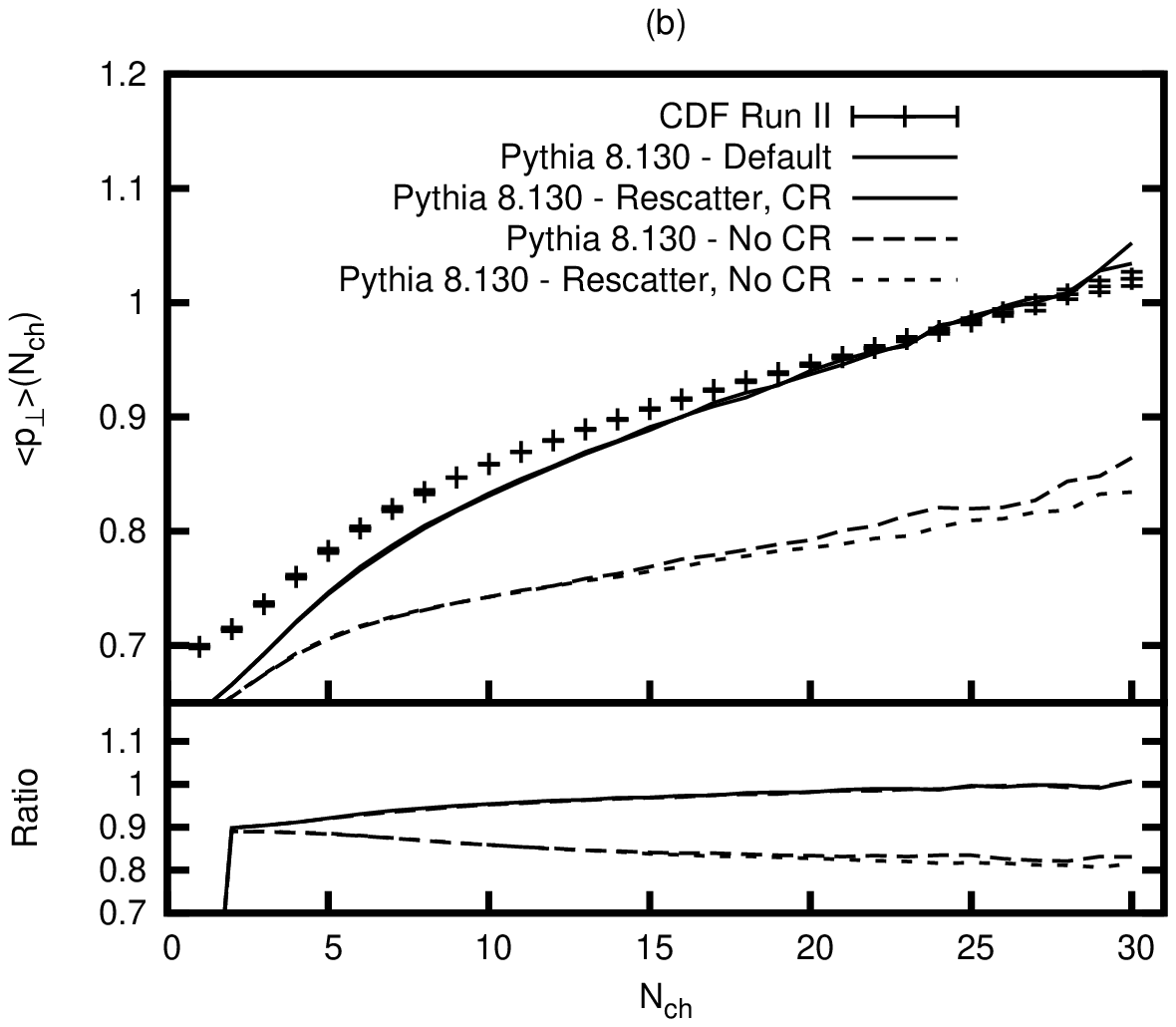}
\end{minipage}
\caption{Mean $\pT$ as a function of charged multiplicity ($\p\pbar$,
$\sqrt{s} = 1.96 \TeV$, $\pT \geq 0.4 \GeV$, $| \eta | \leq 1$). CDF Run
II data is compared to (a) \textsc{Pythia 6}, old and new MPI models and
(b) \textsc{Pythia 8}, new tune, now with rescattering included. With
\textsc{Pythia 8}, the curves with colour reconnection, with and without
rescattering, lie on top of each other
\label{fig:CR}
}
\end{figure}

We next turn our attention to colour reconnection. The key observable is
the mean $\pT$ as a function of the charged multiplicity.
Fig.~\ref{fig:CR}a shows the CDF Run II data \cite{Aaltonen:2009ne} and the
results from the \textsc{Pythia 6} generator. Events are generated using
the soft QCD option and all charged particles with
$\pT \geq 0.4 \GeV$ and $| \eta | \leq 1$
are included, as in the experimental data. With
only soft QCD allowed, the low multiplicity bins may miss some activity,
e.g. from double diffractive events, but it is the higher multiplicities
that concern us here. Both the old (Tune A) and new (Professor
``Pro-pT0'' tune) MPI models in \textsc{Pythia 6} are shown and 
both do well in describing the data. Also shown is the new
model with colour reconnection switched off. Turning reconnection off will
result in an increase in multiplicity, so the procedure used here is to
retune the $\pTo$ parameter of the MPI framework such that the average
charged multiplicity (using the same cuts as above) is the same as when
colour reconnections are switched on. This is a crude retuning method, but
serves well enough to show the desired effects. Without reconnection, and
with a retuned $\pTo$, it is clear that the rise in mean $\pT$ with
multiplicity is much slower and does not come close to reproducing the
data.

Fig.~\ref{fig:CR}b shows the results for \textsc{Pythia 8} (new tune), this
time also with the addition of rescattering. Again, a tuning is performed
such that the average charged multiplicity in each case is the same as the
default curve (some indicative $\pTo$ values, also for LHC energies, are
given in the following section).  The results, although not
completely unexpected, are disappointing. The lack of change when
rescattering is switched on means that the ratio of extra $\pT$ to
multiplicity is essentially unchanged.

As it stands, rescattering does not contribute anything towards
reducing the amount of colour reconnections required to match data.
One point worth considering, however, is that the colour reconnection
algorithm, as described in Sec.~\ref{sec:BR}, has been essentially left
unchanged with the addition of rescattering. One could ask e.g. if 
there are aspects of rescattering systems which means they should be
reconnected in a different way? As such a study could potentially fill
another article, we ignore such questions here, and note that the current
procedure is correct, insofar as colour reconnection is a somewhat
``ad-hoc'' procedure.

\subsection{Cronin Effect}
\label{sec:resultsCronin}
As discussed in Sec.~\ref{sec:MPIintro}, the Cronin effect is the
observation of an enhancement in high-$\pT$ particle production, relative
to low-$\pT$ production, in proton-nucleus collisions compared to
proton-nucleon (after some model-dependent scaling by a nuclear modification
factor). It is postulated that some kind of rescattering contributes to
this effect, so our goal is to see if we can observe any similar type of
effect in $\p\p/\p\pbar$ interactions within our model. Although we are looking at
high-$\pT$ hadron production, the underlying cause within the model is
the (repeated) rescattering of existing partons out to higher overall
$\pT$, even when each rescattering is likely to be rather soft.

The $\pT$-spectra of charged hadrons and neutral pions are taken in the
range $| \eta | < 1.0$ with $\pT > 0.4 \GeV$ using the minimum bias soft QCD
option of \textsc{Pythia}. As before, a retuning of the $\pTo$ parameter of
MPI is performed when rescattering is switched on such that the average
charged multiplicity, using the same cuts as above, is maintained. To give
an idea of the order of magnitude of the changes required to $\pTo$, the
original and retuned values are given in Table~\ref{tab:pTo} for the
different energies and tunes used in this study.

\renewcommand{\arraystretch}{1.15}
\begin{table}
\begin{center}
\begin{tabular}{c|l|c|p{1.5cm}|p{1.5cm}|p{1.5cm}|p{1.5cm}|}
\cline{4-7}
\multicolumn{3}{l|}{}
& \multicolumn{2}{|c|}{\textbf{Old Tune}}
& \multicolumn{2}{|c|}{\textbf{New Tune}} \\
\cline{3-7}
\multicolumn{2}{l|}{}
& $\mathbf{\sqrt{s}~\mathrm{\mathbf{(TeV)}}}$
& \centering $\mathbf{{{\pT}_{0}^{ref}}}$
& \centering $\mathbf{\pTo}$
& \centering $\mathbf{{{\pT}_{0}^{ref}}}$
& \centering $\mathbf{\pTo}$ \tabularnewline 
\cline{2-7}
\multirow{3}{*}{\textbf{Original}\quad}
& \textbf{Tevatron} & 1.96 &
\centering 2.15 & \centering 2.18 &
\centering 2.25 & \centering 2.30 \tabularnewline
& \textbf{LHC}      & 7.00 &
\centering 2.15 & \centering 2.67 &
\centering 2.25 & \centering 3.12 \tabularnewline
& \textbf{LHC}      & 14.0 &
\centering 2.15 & \centering 2.99 &
\centering 2.25 & \centering 3.68 \tabularnewline
\cline{2-7}
\multirow{3}{*}{\textbf{Retuned}\quad}
& \textbf{Tevatron} & 1.96 &
\centering 2.25 & \centering 2.28 &
\centering 2.34 & \centering 2.39 \tabularnewline
& \textbf{LHC}      & 7.00 &
\centering 2.28 & \centering 2.83 &
\centering 2.34 & \centering 3.23 \tabularnewline
& \textbf{LHC}      & 14.0 &
\centering 2.29 & \centering 3.18 &
\centering 2.34 & \centering 3.82 \tabularnewline
\cline{2-7}
\end{tabular}
\end{center}
\caption{$\pTo$ values before and after tuning for Tevatron and LHC
energies with the old and new tunes. The relation between
${\pT}_{0}^{\mathrm{ref}}$ and $\pTo$ is given in
eq.~(\ref{eqn:pT0scaling}). $E_{\mathrm{CM}}^{\mathrm{ref}} = 1800 \GeV$
for both tunes, while $E_{\mathrm{CM}}^{\mathrm{pow}} = 0.16$ for the old
tune and $E_{\mathrm{CM}}^{\mathrm{pow}} = 0.24$ for the new
}
\label{tab:pTo}
\end{table}
\renewcommand{\arraystretch}{1}

\begin{figure}
\centering 
\includegraphics[angle=270,scale=0.65]{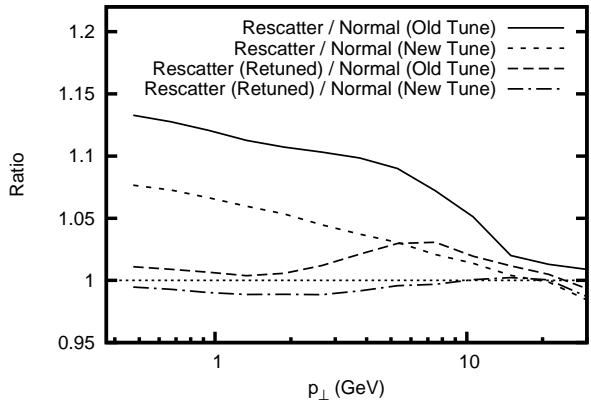}
\caption{Ratios of rescattering (both tuned and untuned) to no rescattering
for parton level inclusive jet distributions, LHC minimum bias
events ($\p\p$, $\sqrt{s} = 14 \TeV$, $\pT > 0.4 \GeV$, $| \eta | < 1.0$,
old and new tunes)
\label{fig:CroninParton}
}
\end{figure}

We begin at the parton level; for an increased hadron production at
high $\pT$, we expect the parton-level $\pT$ distribution to also be
harder. To observe this, we study the parton level inclusive jet distribution
(with ISR, FSR and MPI enabled), where partons must meet the same cuts
described previously. Fig.~\ref{fig:CroninParton} shows the ratio of
the distributions with rescattering switched on (both untuned and tuned)
to the default curve for LHC minimum bias events ($\p\p$, $\sqrt{s} = 14 \TeV$,
old and new tunes). When rescattering is switched on, but no retuning is
performed, there is an overall rise in jet multiplicity across the entire
$\pT$ range, although mostly at low $\pT$, as expected. The retuning is now
performed (using the $\pTo$ values from the hadron level tuning as
described above), such that the excess multiplicity is now removed.

For the new tune, the retuned curve now sits below the default one over
most of the $\pT$ range. Instead, for the old tune, we do indeed have an
excess of partons in the upper-$\pT$ region after retuning. Towards the
highest-$\pT$ scales, the ratios begin to return to unity; a
result of the convolution of falling power-like ($\sim 1 / \pT^4$)
distributions. It is this excess of high-$\pT$ partons which
we would expect to contribute to any Cronin type effect at the hadron
level. The results for other energies and tunes are similar; it is only
with the old tune where there is any visible enhancement in high-$\pT$ jet
production.

\begin{figure}
\begin{minipage}[c]{0.5\linewidth}
\centering 
\includegraphics[angle=270,scale=0.60]{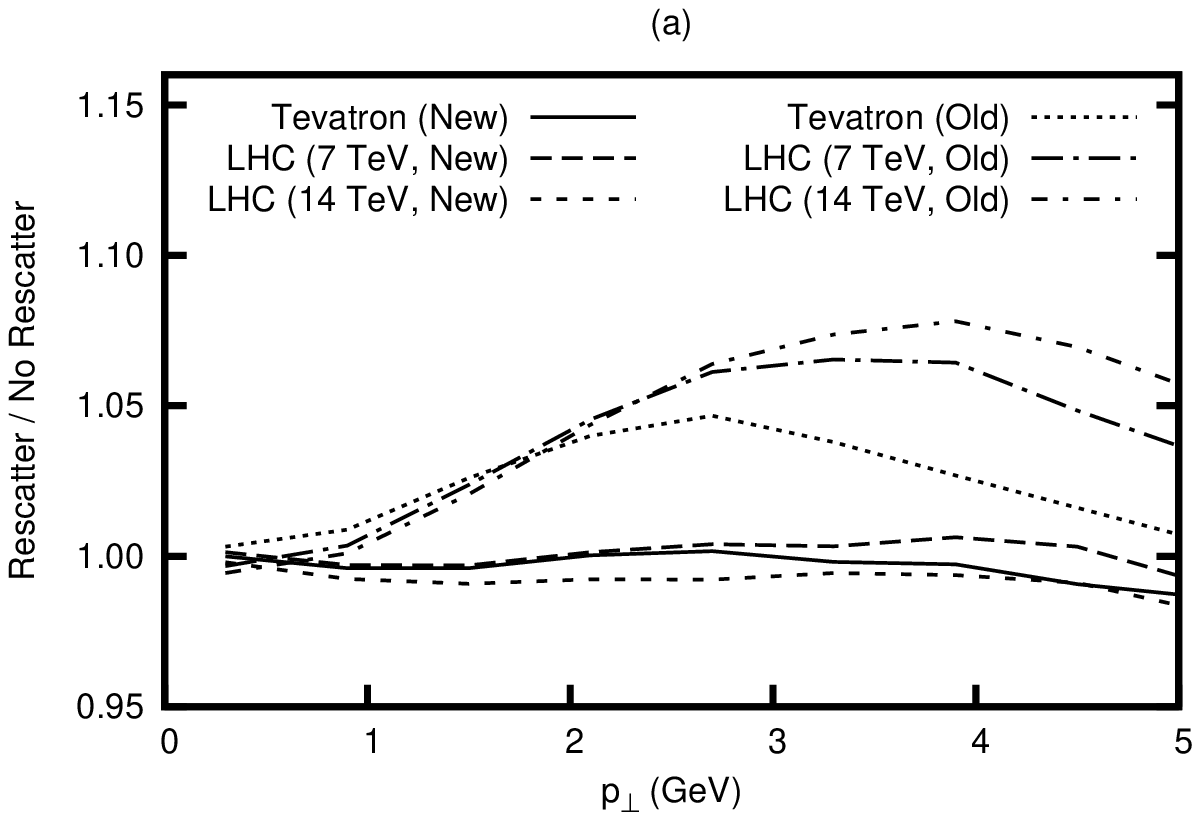}
\end{minipage}
\begin{minipage}[c]{0.5\linewidth}
\centering 
\includegraphics[angle=270,scale=0.60]{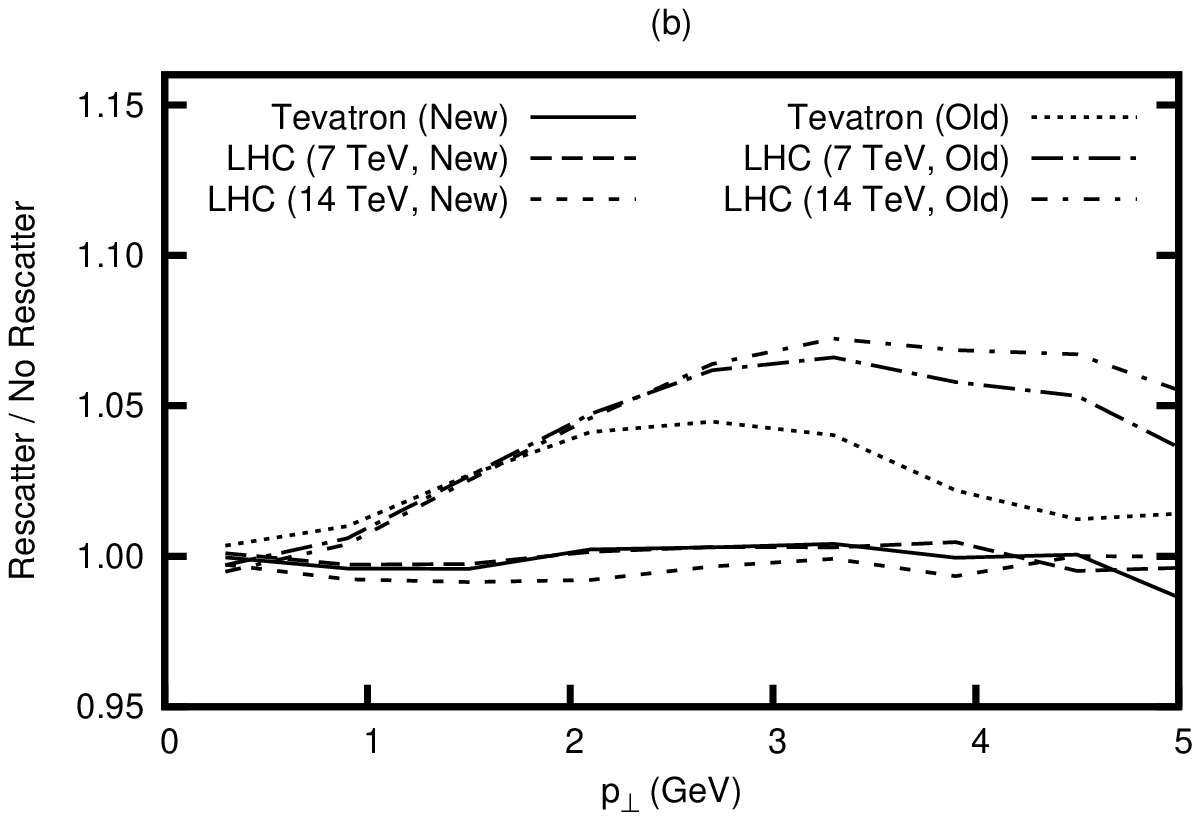}
\end{minipage}
\caption{Ratio of rescattering to no rescattering for the production of (a)
charged hadrons and (b) neutral pions for Tevatron and LHC minimum bias events
($\pT > 0.4 \GeV$, $| \eta | < 1.0$, old and new tunes)
\label{fig:Cronin}
}
\end{figure}

Fig.~\ref{fig:Cronin} now shows the ratio of production with
rescattering enabled to without rescattering for (a) charged hadrons and (b)
neutral pions, for a range of different energies, both with the old and new
tunes. As expected from the parton level study, it is only with the old
tune that there are visible changes. The energy dependence of the different
curves reflects the increased phase space for MPI (and therefore
rescattering) at higher energies.

Most surprising is the lack of enhancement when the new tune is used.
At Tevatron energies, this is not an obvious result, as this is the source
of a large amount of the data used to tune the MPI model. This shows, then,
one of the difficulties in tuning an event generator; contributions can be
shifted around between the different models through parameter adjustments
(in this case a change in the hadronic matter profile and a slight change
in $\pTo$ and other parameters), such that the same data can be described
in different ways.  We have also seen that the underlying cause is the lack
of high-$\pT$ enhancement at the parton level, but an interesting question
is how, after retuning, the overall lower multiplicity at the parton level
(Fig.~\ref{fig:CroninParton}) leads to the same average charged
multiplicity at the hadron level. Most likely it is related to the more
complicated colour flows that result from rescattering. This then ties in
with the issue of colour reconnections, as discussed in the previous
section, and as such we do not currently study this further.

\subsection{Jet Study}
\label{sec:jetstudy}
Finally, we turn to a ``jet'' study of rescattering.  We feed the
hadron-level results of \textsc{Pythia} ($| \eta | < 3.5$, $\pT > 0.4 \GeV$)
into FastJet and use the anti-$\kT$ algorithm ($R = 0.4$)
\cite{Cacciari:2005hq,*Cacciari:2008gp}. When rescattering is enabled, the
same $\pTo$ values as Sec.~\ref{sec:resultsCronin} are used for simplicity.

\begin{figure}
\begin{minipage}[c]{0.5\linewidth}
\centering 
\includegraphics[angle=270,scale=0.60]{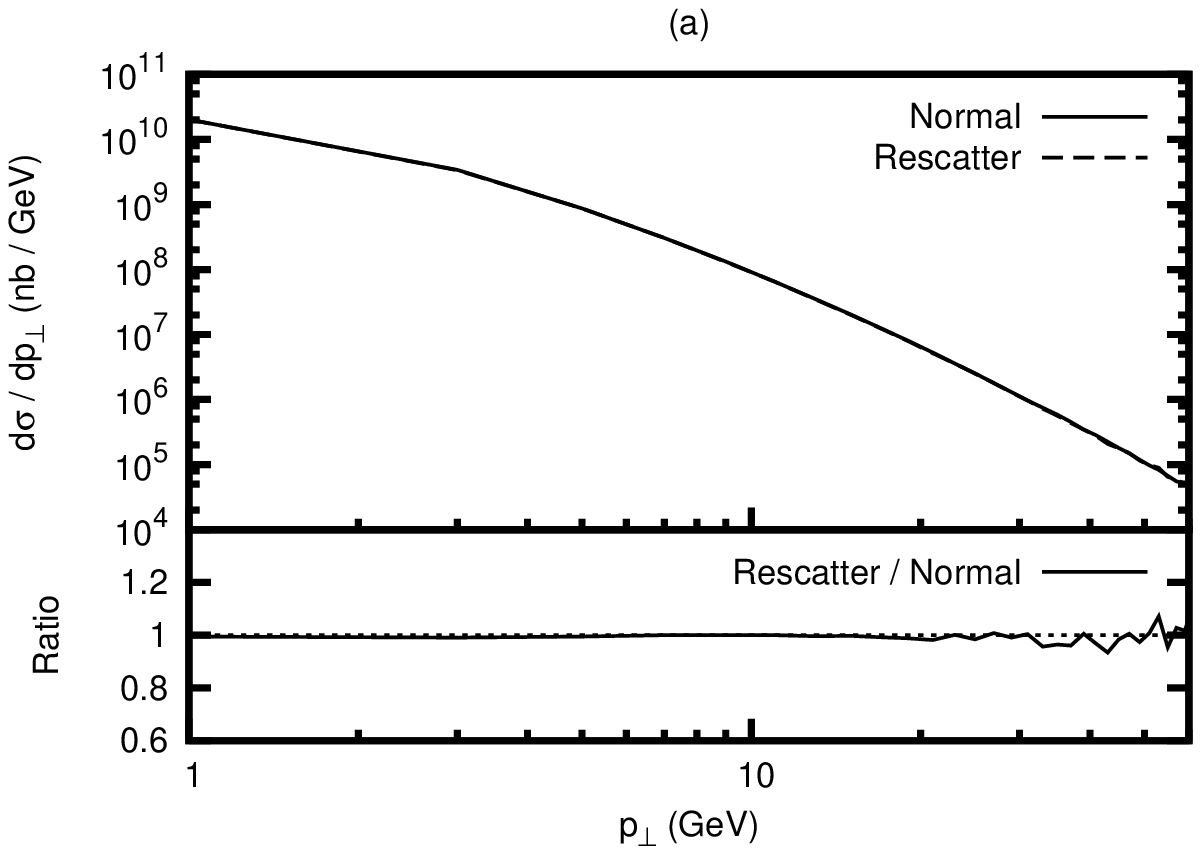}
\end{minipage}
\begin{minipage}[c]{0.5\linewidth}
\centering 
\includegraphics[angle=270,scale=0.60]{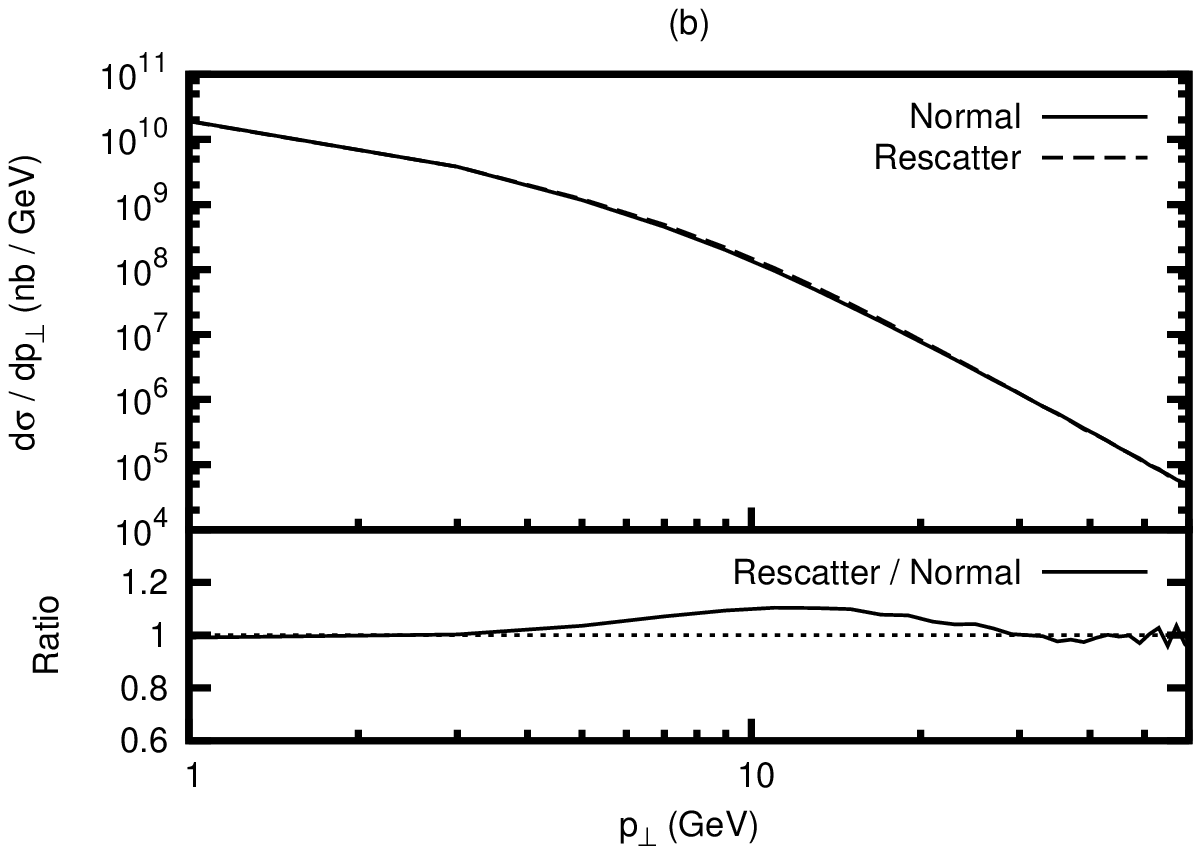}
\end{minipage}
\caption{Inclusive jet $\pT$ distribution for LHC minimum bias events 
($\p\p$, $\sqrt{s} = 14 \TeV$), (a) new tune, (b) old tune. The results
with rescattering switched on are almost indistinguishable from the normal
curves, but with the old tune the ratio does show a slight excess
\label{fig:JetInc}
}
\end{figure}

The starting point is Fig.~\ref{fig:CroninParton}; at the parton level,
with the old tune, we have seen an excess of high-$\pT$ partons, so we aim
to see if this is also true after hadronisation. In Fig.~\ref{fig:JetInc},
the inclusive jet distribution is shown for those jets with $| \eta | <
1.0$ for LHC minimum bias events ($\p\p$, $\sqrt{s} = 14 \TeV$, new and old
tunes). The results follow the expected pattern; with the old tune, there is
an excess of higher-$\pT$ jets, while with the new tune, there is almost no
change.

\begin{figure}
\centering 
\includegraphics[angle=270,scale=0.60]{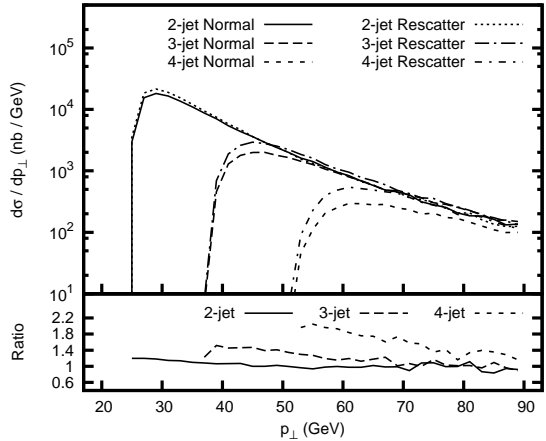}
\caption{Two-, three- and four-jet exclusive cross sections for LHC minimum bias
events ($\p\p$, $\sqrt{s} = 14 \TeV$, ${\pT}_{\mrm{jet}} > 12.5 \GeV$,
$|\eta| < 1.0$, old tune)
\label{fig:JetExc}
}
\end{figure}

We now extract the two-, three- and four-jet exclusive cross sections, where all jets
must lie within $|\eta| < 1.0$ and have a minimum transverse momentum
${\pT}_{\mrm{jet}} > 12.5 \GeV$. From Fig.~\ref{fig:JetInc}, we see that it is with
this $\pT$ cut that we expect the greatest rise in the different jet rates,
and this is shown in Fig.~\ref{fig:JetExc} for the old tune as a function
of the summed $\pT$ of the jets. There is a clear rise in the jet cross
sections. Although, in itself, this is not any kind of definitive signature
of rescattering, it is still encouraging. Rescattering does produce
noticeable effects which will affect e.g. a full tuning of the generator.

As a final step, we check for kinematical signatures of rescattering.
With DPS, for example, the different pairwise interactions are clearly
distinct and azimuthal distributions can be used as a distinguishing
signature. Rescatterings are not distinct in this way. They are very deeply
entangled in the downward evolution of the final state, so it is not clear
if any such signatures may exist. For the three-jet sample of
Fig.~\ref{fig:JetExc}, we study the smallest $\Delta R$ value between the
different pairs of jets per event. One could hope that the distribution of
$\Delta R$ values from three-jet events where rescattering is involved is
somehow different than the background events (e.g. three-jet events from
radiation which may be characteristically peaked in the small $\Delta R$
region). For the four-jet sample, we instead study the smallest and largest
$\Delta \phi$ values between the jets per event. It is known that DPS
events have a characteristic $\Delta \phi$ peak at $\pi$, but there are
also radiative contributions which can mask rescattering. Unfortunately,
for both samples, the results with and without rescattering are essentially
indistinguishable.

\section{Conclusions}
\label{sec:conclusions}

In this article we have presented a model for rescattering in MPI,
allowing the full generation of events from a central simple process
to the multiparticle hadron-level final state. To the best of our
knowledge, this is the first time that rescattering has been modeled
in such a detailed manner. The model is implemented and available for
public use inside the \textsc{Pythia 8} event generator. The
model---generator connection is very important here; MPI physics is
so complicated, and hovering so near to the brink of nonperturbative
physics, that purely analytical approaches have a limited range of
validity.

The formalism outlined in Sec.~\ref{sec:rescatteringPy8} provides a
method of including already scattered partons back into the PDFs of the
hadron beams such that they can be rescattered. Further, in
Sec.~\ref{sec:beamassociation}, we have shown that a natural kinematical
suppression means that the importance of the different beam association
procedures is reduced. The main technical challenges, then, come with the
inclusion of radiation and beam remnants. The kinematics of a dipole-based
parton shower, combined with the flow of colour from one scattering
subsystem into another, can lead to potentially large momentum imbalances
in those stages of event generation that use rotations and Lorentz boosts
to adjust parton kinematics (namely ISR and primordial $\kT$). We have
found that the ``trick'' of shifting the momenta of internal lines, to
always conserve system momentum, does not have large adverse effects on
event generation and allows us to integrate rescattering fully into the
existing framework.
 
As we have seen, rescattering is a common occurrence. At the LHC we
predict that of the order of half of all minimum-bias events will
contain at least one rescattering, and for events with hard processes
the fraction is even larger. Nevertheless, evaluating the effects of
rescattering is challenging since it is a secondary effect within
the MPI machinery. The precise amount of MPI has to be tuned to data,
e.g. by varying the $\pTo$ turn-off parameter, so the introduction of
rescattering to a large extent can be compensated by a slight decrease
in the amount of ``normal'' $2 \to 2$ MPIs. Worse still, most MPIs are
soft to begin with, and rescattering then introduces a second scale,
even softer than the ``original'' scattering. Thus, rescattering is typically
associated with particle production at the lower limit of what can be
reliably detected. We have shown some effects of the Cronin type in
hadronic $\pT$ spectra, but too small to offer a convincing signal.

If one zooms in on the tail of events at larger scales, the rate of
(semi-)hard three-jet events from rescattering is of the same order in
$\alphas$ as four-jets from DPS, but the background from initial- and
final-state radiation starts one order earlier. Furthermore, DPS
has some obvious characteristics to distinguish it from $2 \to 4$
radiation topologies: pairwise balanced jets with an isotropic relative
azimuthal separation. No corresponding unique kinematical features are
expected for rescattering and we have not found any new ones.

To some extent, this is a disappointing outcome of a project that has
required the lengthy development of a complex, sophisticated machinery.
However, let there be no doubt; rescattering is a logical consequence
of MPI. In recent years, MPI has gained a wide acceptance, and come to
be one of the cornerstones in our picture of hadronic physics. Therefore
we can no longer rest content with qualitative estimates and models;
we have to come up with quantitative answers to a number of detailed
questions. The role of rescattering is one such, and certainly not the
last one to need detailed modeling.

We should also not exclude that experimentalists may come up with new
observables where rescattering is better visible, or tensions in the
data that find an explanation with the presence of rescattering. Now that
we can supply a complete implementation of rescattering, at least
experimental studies will not be restricted by the theory support
in this area. The next logical step, then, would be to produce full
\textsc{Pythia 8} tunes, with and without rescattering switched on
(our procedure of only raising the $\pTo$ parameter when rescattering
is included is certainly too crude).  Effects can then be studied
across the range of observables used in tuning the generator, and
changes to the parameters for the model with rescattering will give
an insight into the role that it plays.

In the future we also hope to further improve our modeling.
There are other areas left to address within the context of
detailed models for MPI, such as the correlation between momentum
fraction and transverse spread of partons in a hadron. Also
areas not directly related to the MPI framework as such can be
important, such as the colour reconnection procedure. There is such
a hint of potential shortcomings in Sec.~\ref{sec:resultsCronin},
where, with rescattering and the new tune, the parton-level multiplicity
drops, and yet the hadron-level multiplicity stays the same.

So it is clear that this study is not the end of the story. Once the LHC
starts, we can look forward to a wide range of new data on minimum-bias
and underlying-event physics, which may further stimulate us to refine
and rethink the MPI models used in current event generators.
With rescattering, we have presented one possible future direction
for models and studies, which we hope can contribute towards a greater
understanding of MPI at current and future colliders.

\subsection*{Acknowledgments}

The authors wish to thank Florian Bechtel for his involvement in this
project in an early phase.

This work was supported in by the Marie Curie Early Stage Training program
``HEP-EST'' (contract number MEST-CT-2005-019626) and in part by the Marie
Curie research training network ``MCnet'' (contract number
MRTN-CT-2006-035606).

For Feynman diagrams, Jaxodraw \cite{Binosi:2003yf} was used.

\bibliography{rescatter}{}
\bibliographystyle{utcaps}

\end{document}